%% file: ms.tex
\documentclass[useAMS,usenatbib]{mn2e}
\usepackage{graphicx}
\usepackage{lscape}

\title[A deep WISE search for very late type objects]{A deep WISE search for very late type objects and the discovery of two halo/thick-disk T dwarfs: WISE 0013+0634 and WISE 0833+0052}
\author[D. J. Pinfield et~al.]{D. J. Pinfield$^{1}$\thanks{E-mail: D.J.Pinfield@herts.ac.uk}, 
J. Gomes$^{1}$, A. C. Day-Jones$^{2,1}$, S. K. Leggett$^{3}$, M. Gromadzki$^{4}$, 
\newauthor
B. Burningham$^{1}$, M. T. Ruiz$^{2}$, R. Kurtev$^{4}$, T. Cattermole$^{1}$, C. Cardoso$^{5}$, N. Lodieu$^{6,7}$, 
\newauthor
J. Faherty$^{2,8}$, S. Littlefair$^{9}$, R. Smart$^{5}$, M. Irwin$^{10}$, J. R. A. Clarke$^{1}$, L. Smith$^{1}$, 
\newauthor
P. W. Lucas$^{1}$, M. C. G\'{a}lvez-Ortiz$^{11}$, J. S. Jenkins$^{2}$, H. R. A. Jones$^{1}$, R. Rebolo$^{6}$, 
\newauthor
V. J. S. B\'{e}jar$^{6,7}$, B. Gauza$^{6,7}$
\\
$^{1}$Centre for Astrophysics Research, Science and Technology Research Institute, University of Hertfordshire, Hatfield AL10 9AB, UK \\
$^{2}$Universidad de Chile, Santiago, Casilla 36-D, Chile \\
$^{3}$Gemini Observatory, Northern Operations Center, 670 N. A'ohoku Place, Hilo, HI 96720, USA \\
$^{4}$Departamento de F\'{i}sica y Astronom\'{i}a Universidad de Valpara\'{i}so, Av. Gran Breta\~{n}a 1111, Playa Ancha, Casilla 5030, Chile \\
$^{5}$Istituto Nazionale di Astrofisica, Osservatorio Astronomico di Torino, Strada Osservatrio 20, 10025 Pino Torinese, Italy \\
$^{6}$Instituto de Astrof\'isica de Canarias, 38200 La Laguna, Spain \\
$^{7}$Dept. Astrof\'isica, Universidad de La Laguna, E-38206 La Laguna, Tenerife, Spain \\
$^{8}$Department of Astrophysics, American Museum of Natural History, Central Park West at 79th Street, New York, NY 10034, USA \\
$^{9}$Dept of Physics and Astronomy, University of Sheffield, Sheffield S3 7RH, UK \\
$^{10}$Institute of Astronomy, Madingley Road, Cambridge CB3 0HA, UK \\
$^{11}$Centro de Astrobiolog\'ia (CSIC-INTA), Ctra. Ajalvir km 4, E-28850 Torrej\'on de Ardoz, Madrid, Spain
}

\begin{document}
\include{aas_macros}

\maketitle

\begin{abstract}
A method is defined for identifying late T and Y dwarfs in WISE down to low values of signal-to-noise. This requires a WISE detection only in the $W2$-band and uses the statistical properties of the WISE multi-frame measurements and profile fit photometry to reject contamination resulting from non-point-like objects, variables and moving sources. To trace our desired parameter space we use a control sample of isolated non-moving non-variable point sources from the SDSS, and identify a sample of 158 WISE $W2$-only candidates down to a signal-to-noise limit of 8. For signal-to-noise ranges $>$10 and 8-10 respectively, $\sim$45\% and $\sim$90\% of our sample fall outside the selection criteria published by the WISE team (Kirkpatrick et al. 2012), due mainly to the type of constraints placed on the number of individual $W2$ detections. We present follow-up of eight candidates and identify WISE 0013+0634 and WISE 0833+0052, T8 and T9 dwarfs with high proper motion ($\sim$1.3 and $\sim$1.8 arcsec~yr$^{-1}$). Both objects show a mid-infrared/near-infrared excess of $\sim$1-1.5 magnitudes, and are $K-$band suppressed. Distance estimates lead to space motion constraints that suggest halo (or at least thick disk) kinematics. We then assess the reduced proper motion diagram of WISE ultracool dwarfs, which suggests that late T and Y dwarfs may have a higher thick-disk/halo population fraction than earlier objects.
\end{abstract}


\section{Introduction}
\label{sec:intro}


Population studies of brown dwarfs \citep[mass$<$0.075M$_{\odot}$;][]{baraffe2003} can provide crucial constraints on the low mass extreme of the star formation process \citep[e.g.][]{reipurth2001,bate2002,goodwin2007,bonnell2008,stamatellos2009,machida2009}, through measurements of the luminosity and mass functions \citep[e.g.][]{cruz2007,metchev2008,pinfield2008,burningham2010a,reyle2010,bastian2010,kirkpatrick2012,lodieu2012b,burningham2013}, and sampling of the formation history \citep[e.g.][]{dayjones2013}. Also the cooling of brown dwarfs leads to a $T_{\rm eff}$ range that encompasses the lowest mass stars and giant planets. As such brown dwarfs make excellent test beds to improve our understanding of sub-stellar atmospheres \citep[e.g.][]{pinfield2006,burningham2009,king2010,zhang2010,dayjones2011,pinfield2012,gomes2013,faherty2012b}.

Brown dwarf discovery began in 1995 \citep{rebolo1995,nakajima1995}, and has expanded, mainly through the use of large scale surveys, into several new spectral classes. The 2MASS \citep{skrutskie2006}, DENIS \citep{epchtein1999}, and SDSS \citep{york2000} surveys revealed the dusty L dwarfs \citep[with $T_{\rm eff}$=2300-1400K; e.g.][]{kirkpatrick1999}, the methane dominated T dwarfs \citep[with $T_{\rm eff}<$1400K; e.g.][]{strauss1999,geballe2002,burgasser2006}, and the LT transition objects that join together these two spectral types \citep[e.g. ][]{leggett2000}. More recently the UKIDSS surveys and the Canada France Brown Dwarf Survey have identified the coolest examples of the T dwarf class in the 500-700K range \citep[e.g.][]{warren2007,burningham2008,delorme2008,burningham2010b,lucas2010}, with large scale deep near-infrared coverage. However, it took the mid-IR sensitivity of the WISE observatory \citep{wright2010} to break through into the $<$500K regime for free-floating field objects. Although WISE has a similar sensitivity to L dwarfs as 2MASS \citep[thus providing good scope for infrared proper motion surveys;][]{castro2011,gizis2011a,gizis2011b,castro2012}, the relative dominance of mid-infrared emission for $<$500K objects \citep[particular around 4.5 microns;][]{burrows2003,mainzer2011,allard2012} leads to a greatly increased sensitivity at this $T_{\rm eff}$.

Such objects are known as Y dwarfs. Like late T dwarfs, Y dwarf spectra exhibit deep H$_2$O and CH$_4$ absorption bands which dominate their spectral morphology. However, the Y dwarf WISE 1828+2650 \citep[][; presented as the archetype]{cushing2011} has notably extreme colours ($J-W2$=9.3) and a spectrum that is markedly different to the latest T dwarfs \citep[e.g. UGPS 0722-0540; ][]{lucas2010}. The $J-H$ colour of WISE 1828+2650 is much redder than those of the late Ts showing that a reversal of the previously known colour trend has occurred, indicative of a collapse in the near-infrared flux relative to that at $\sim$5 microns \citep{cushing2011}. Furthermore, the [3.6]-[4.5] Spitzer colour shows a turn to the blue that may be indicative of a shift in position of the $\sim$5 micron flux peak \citep[][; K12 hereafter]{kirkpatrick2012}. WISE 1828+2650 would thus seem to fulfill at least one of the five criteria put forward by \citet{burrows2003} for the establishment of the Y class. Despite the marked difference between the late T dwarfs and the archetypal Y dwarf, the actual transition follows a gradual change (as per the norm) which is predominantly quantified by the narrowness of the $J$-band flux peak \citep[see also][]{mace2013}.

By comparison to WISE, the UKIDSS Large Area Survey and Galactic Plane Survey cover $\sim$7000 sq degs of sky down to $J\sim$19.5 \citep{lawrence2007}, with VISTA now expanding into the southern sky \citep[e.g.][]{lodieu2012a,gauza2012}. However, Y dwarfs generally have $J-W2\ge$5, and the first WISE Y dwarfs were found in about half the sky in the $W2$=14-14.5 range \citep[][; K11 hereafter]{cushing2011,kirkpatrick2011}. From a Y dwarf perspective the WISE W2 coverage and sensitivity in the first incremental release was equivalent to a near-infrared survey reaching $J>$19.5 over half the sky. Indeed, when considering the reddest Y dwarfs currently known (e.g $J-W2\sim$9) WISE is sensitive out to $\sim$10 times the distance of the deepest near-infrared surveys.

The collaboration led by Kirkpatrick of the WISE Science team is implementing an ongoing search for late T and Y dwarfs in WISE (K11, K12). Though maintaining a variety of candidate lists, their general search method has been characterised in K12. As well as a set of colour, magnitude, and coordinate constraints, and allowances for non-detection in some bands, they placed constraints on a set of criteria parameterising source detection and character. Thresholds were used for the number of detections in individual frames (to remove spurious aligned artefacts), blending with other sources was avoided (since it can result in poorly determined photometry), and the point-like nature of a source was assessed. This programme discovered the first Y dwarfs \citep{cushing2011}, and has to-date identified a population of fourteen objects in this new class \citep[K12, ][]{tinney2012}.

In another WISE search \citet{pinfield2012} searched for wide companion late T and Y dwarfs around Hipparcos and Gliese stars. This search had similar colour constraints and allowances for non-detection in various bands, but did not place any further constraints on source detection and character. This was a practical approach in this case because only candidate widely separated companions were taken forward for closer inspection, and the sky coverage was severely limited by the requirement for proximity to Hipparcos and Gliese stars. The distances of the possible primary stars also contributed to the assessment (since companions would be essentially at the same distance as their primary), and visual inspection could proceed for a few tens of candidates after an automated selection process. This search identified a new metal poor T8p companion, and recovered  three other previously known objects (T6p, T8 and T8.5).

In this paper we describe a new search method designed to effectively identify late T and Y dwarf candidates in the full WISE sky down to a faint detection limit in $W2$. Section \ref{sec:initial_selection}-\ref{sec:rejection_methods} presents our selection and rejection method, making use of a control sample to assess $W2$ source characteristics. We initially search down to $W2$ signal-to-noise of 10, and Section \ref{sec:extending_search} extends our search down to a signal-to-noise of 8. We discuss our visual inspection process in Section \ref{sec:vis_inspect}, and the new candidate sample itself in Section \ref{sec:new_sample}. Section \ref{sec:classification} describes a classification system that allows us to prioritise candidates for further follow-up, and Section \ref{sec:phot} describes our current photometric followup measurements. Our analysis of the recently identified WISE 1639-6847 is presented in Section \ref{sec:wise1639}. And the two newly discovered objects WISE 0013+0634 and WISE 0833+0052 are discussed in Section \ref{sec:wise0013_0833}. We plot and assess these new objects, and also other L T and Y dwarfs with measured proper motion, in reduced proper motion diagrams in Section \ref{sec:rpmd}. Future work is considered in Section \ref{sec:sum}.


\section{Candidate selection}
\label{sec:selection}

\subsection{The initial selection}
\label{sec:initial_selection}

In typical WISE sky (10-30 exposures) the 10-$\sigma$ limit occurs at W2=15-15.7, and the detection limits (2-$\sigma$) in the other bands are W1$\simeq$17.6, W3$\simeq$12.3 and W4$\simeq$9. WISE Y dwarf colours are $W1-W2>$3.9, $W2-W3$=1.7-2.6 and $W2-W4\simeq$5, so in general we thus expect Y dwarfs around the $W2$ 10-$\sigma$ limit to be non-detections in the other bands. We thus began by selecting All-Sky catalogue sources that were detected with a signal-to-noise of at least ten in the $W2$-band ({\tt{w2snr}}$\ge$10), but that were not detected in the $W1$-,  $W3$- and $W4$-bands ({\tt{w*mpro}} eq null or {\tt{w*sigmpro}} eq null, meaning that the profile-fit magnitude is a 95\% confidence upper limit or that the source is not measurable). We also required that there was no 2MASS source within 3 arcseconds of the WISE source, and at least 8 individual exposures in each bands ({\tt{w*m}}$\ge$8). \citet{wright2010} explain how the WISE mission design provides at least 8 exposures over more than 99\% of the sky during a 6 month mission.

To minimise contamination from reddened objects, we removed sources towards reddened regions of sky. To estimate reddening we used the on-line Galactic Dust Reddening and Extinction Calculator at the NASA/IPAC Infrared Science Archive, which computes line-of-sight inter-stellar extinction using the technique pioneered by \citet{schlegel1998}. To reject stars with reddened $W1-W2>$2.0 we aimed to exclude regions with $E(W1-W2)\ge$1.4 (since normal stars have $W1-W2$=0.0-0.4 and will scatter somewhat in colour due to photometric uncertainty). This constraint was transformed into an $A_v$ limit using the extinction curves of \citet{mathis1990} leading to a rejection criterion where the average line-of-sight visual extinction $A_v>$0.45. We also allowed for extinction variation of 0.35 magnitudes (due to dense structures) across the $\sim$5 arcminute on-sky resolution of the extinction maps (this range of variations is exceeded in only $\sim$1\% of the sky). We thus rejected sources towards regions of sky with average line-of-sight extinction $A_v>$0.8 (21.7\% of the sky).

We experimented with the {\tt{ccflag}} parameter to remove sources flagged as possible spurious detections of diffraction spikes, persistence, halos or optical ghosts in the $W2$-band. However, there was no such contamination in our initial selection, which consisted of 6067 sources with {\tt{w2snr}}$\ge$10.

\subsection{Refining the selection}
\label{sec:refining_selection}

We devised a set of additional selection criteria to remove contamination in the form of (i) resolved (or partially resolved) sources such as galaxies and sources associated with nebulosity or artefacts, (ii) variable sources, and (iii) sources that move significantly over the time-scale of the multi-frame images (i.e. solar system objects).

\subsection{The control sample}
\label{sec:control_sample}

As a means to differentiate between the desired database properties of quality sources and those of contamination that we wish to reject, we defined a control sample of isolated point-like non-variable non-moving sources, extending all the way down to the faint survey limits. To define this control sample we combined the Sloan Digital Sky Survey (SDSS) with WISE. We selected sources spectroscopically classed by SDSS as stars, and to avoid field crowding (and thus prefer isolated sources) we limited our search to a galactic latitude $|b|>$70 deg (also removing giant stars that could be variable). We used photometric limits of $g<$20.0 and $g-r<$0.3 to select stars with reasonably blue colour and WISE magnitudes down to $W2$=18. We selected point-like Sloan sources (that were un-blended in WISE) with proper motions $>$20 mas~yr$^{-1}$ which we cross-matched with WISE using a 1 arcsecond cross-match radius to avoid mis-matches. We thus exclude non moving galaxies and high proper motion objects, with the control sample thus consisting of stellar point sources that are stationary across the WISE multi-frame measurements.

\subsection{Rejection methods}
\label{sec:rejection_methods}

\begin{figure*}
\begin{center}
\includegraphics[height=22.0cm, angle=0]{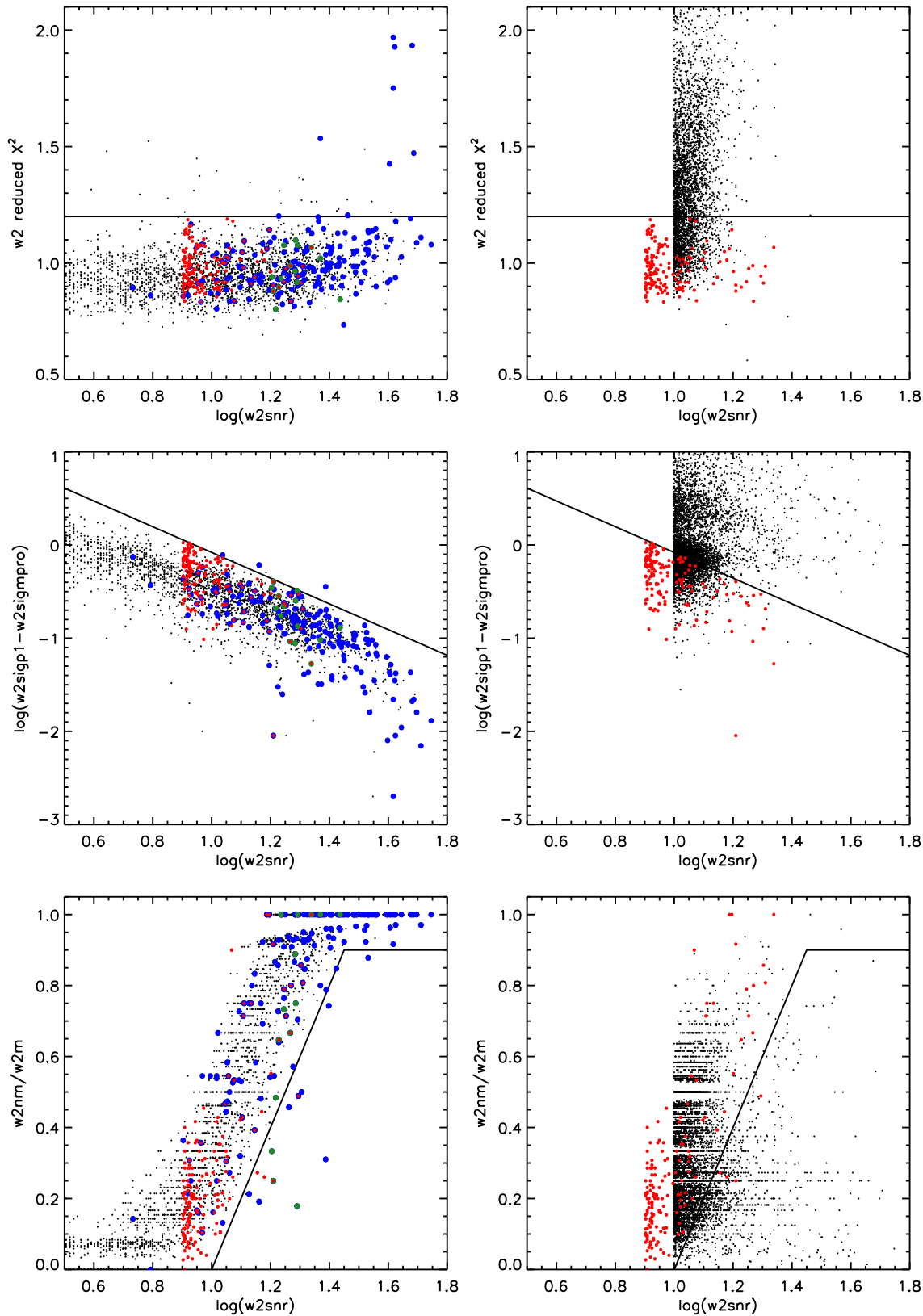}
\caption{The plots show the 3 rejection methods employed by our search, which focus on assessing the profile fit photometry, the photometric uncertainties, and the detection numbers. Plots on the left-hand side show the control sample (see text), as well as the K12 census (blue, except for Y dwarfs which are green). Plots on the right show our initial {\tt{w2snr}}$>$10 selection (small black symbols). Our rejection cuts are indicated with solid lines, and our final candidates are over-plotted as red symbols. \label{fig:selection}}
\end{center}
\end{figure*}

In Figure \ref{fig:selection} the plots on the left hand side show the control sample (small black points), as well as sources from the All-Sky late T and Y dwarf census of K12 as blue symbols, except for Y dwarfs which are green. We plot our final selection of new candidates with {\tt{w2snr}}$>$8 in red (see Section \ref{sec:extending_search} for the {\tt{w2snr}}=8-10 selection). The plots on the right hand side of Figure \ref{fig:selection} show how contamination is rejected from the 6067 $W2$-only initial selection with {\tt{w2snr}}$>$10. The rejection criteria that we employed are indicated with solid lines, and discussed below.

The top plots show our `profile fit photometry rejection method' which assesses how point-like each source is. The {\tt{w2rchi2}} parameter is plotted against log({\tt{w2snr}}), where {\tt{w2snr}} is the signal-to-noise of the $W2$ measurements. {\tt{w2rchi2}} is the reduced $\chi^2$ of the $W2$ profile fit photometry, and is a direct indicator of how well the source is represented by the optimised point-spread function fit. It can be seen from the control sample plot (top left) that in all but a small number of outlier cases ($\sim$1\%), {\tt{w2rchi2}} lies in a band between 0.7 and 1.2 for {\tt{w2snr}} from $2-50$ (log({\tt{w2snr}})=0.3-1.7). Seven of the K12 census ($\sim$4\%) have {\tt{w2rchi2}}$>$1.2, and visual inspection showed that the majority were very close to or slightly overlapping with neighbouring sources in the WISE images (although they are listed as un-blended in the WISE database). In these cases this effect has led to an increased {\tt{w2rchi2}} value for the profile fit photometry. We also note that in a few cases there was no clear explanation for the increased {\tt{w2rchi2}} values, and these sources could warrant further study for unresolved multiplicity \citep[e.g.][]{gelino2011,liu2011}. This rejection method should thus be effective at retaining isolated point sources, but will reject a small fraction where field crowding and blending are an issue.

The middle plots show our `photometric uncertainty rejection method'. We plot log({\tt{w2sigp1}}$-${\tt{w2sigmpro}}) against log({\tt{w2snr}}), where {\tt{w2sigmpro}} is the integrated flux uncertainty and {\tt{w2sigp1}} is the standard deviation of the population of $W2$ fluxes measured on the individual frames (both in magnitude units). For isolated non-variable non-moving objects {\tt{w2sigp1}} provides a direct measure of the uncertainty on individual frames, which should relate to the integrated flux uncertainty. Indeed, the control sample (middle left plot) traces out a well defined sequence over the full signal-to-noise range, though there are two K12 sources ($\sim$1\% of the census) slightly above this sequence. However, visual inspection once again showed that these two sources slightly overlapped with neighbouring sources in the WISE images, which has led to an increase in the {\tt{w2sigp1}} values. Guided by the control sample we rejected all sources that lie above the upper limit line in the plot, log({\tt{w2sigp1}}$-${\tt{w2sigmpro}})$=1.3-1.38$ log({\tt{w2snr}}).

The bottom plots show the `detection number rejection method' that we used when there were a limited number ($<$8) of detections in the individual frames. The plots show the fraction of individual $W2$ frames in which the source was detected with a signal-to-noise greater than three (i.e. {\tt{w2nm}}$/${\tt{w2m}}), against log({\tt{w2snr}}). The control sample traces out a sequence in this parameter-space (bottom left plot), ranging from detection in all frames down to no individual detections (above a signal-to-noise of 3). We thus define a minimum fraction of frames in which the $W2$ source must be detected, shown by the solid line where its sloping portion is defined by {\tt{w2nm}}$/${\tt{w2m}}$<$1.8 log({\tt{w2snr}})$-$1.7. Although eleven of the K12 census (6\%) lie below this line, all eleven are detected in at least eight $W2$ exposures and thus would not be rejected by our method. However, this is an indication of what could be rejected amongst sources detected in $<$8 frames. Visual inspection showed that these outliers were all close to or overlapped with neighbouring sources in the WISE images, which clearly affects the ability of the WISE source detection algorithm to identify the source in individual frames. We therefore expect a low level of incompleteness (mostly in crowded fields) due to this rejection method. We rejected sources that were detected in less than 8 frames if they lay below the rejection line.

Table \ref{tab:search} summarises the number of {\tt{w2snr}}$>$10 candidates that remain after rejection using our three different methods. Results are shown for different orders of applying the methods, to indicate their effectiveness and highlight common ground amongst the rejection methods. It can be seen that the profile fit photometry method rejects the most objects, followed by the photometric uncertainty rejection method. It is also clear that the profile fit photometry method rejects most of the sources rejected by the detection number method. So the overall approach might be simplified without substantially reducing the number of rejected sources, by omitting the detection number method. However, we chose to apply all three methods here, rejecting 5163 sources and leaving 904 {\tt{w2snr}}$>$10 sources for visual inspection.

\begin{table}
\begin{center}
\begin{tabular}{|c|c|c|c|}
\hline
\multicolumn{4}{|c|}{{\tt{w2snr}}$\ge$10 selection}\\
\multicolumn{4}{|c|}{Initial selection: 6067}\\
\hline
\multicolumn{4}{|c|}{Applied 1st}\\
        & psf  & sig  & det  \\
        & 1394 & 2804 & 5168 \\
\hline
\multicolumn{4}{|c|}{Applied 2nd}\\
psf     & --   & 923  & 1342 \\
sig     & 923  & --   & 2483 \\
det     & 1324 & 2483 & --   \\
\hline
\multicolumn{4}{|l|}{After psf/sig/det rejections: 904}\\
\hline
\end{tabular}
\end{center}
\caption{Summary of how the three main rejection methods incrementally reduce the initial {\tt{w2snr}}$\ge$10 selection. In the table, $psf$ refers to the profile fit photometry rejection method, $sig$ to the photometric uncertainty rejection method, and $det$ to the detection number rejection method.\label{tab:search}}
\end{table}

\subsection{Extending the search to the 8-$\sigma$ $W2$ limit}
\label{sec:extending_search}

We built on our {\tt{w2snr}}$>$10 selection by extending our analysis into the {\tt{w2snr}}=8-10 range. We made an initial selection of 18,112 sources, and our three rejection methods (from Section 2.4) left 3252 sources. A visual inspection of $\sim$100 of these revealed that many were found close to bright stars, and resulted from spurious or poor quality detections within the extended bright star halos. We therefore implemented an additional rejection method for this {\tt{w2snr}} range, to remove such contamination prior to visual inspection. The contamination and confusion flag ({\tt{cc\_flags}}) parameter in the All-Sky Database assesses such contamination, though we chose to establish our own halo criterion so as to optimize this rejection method for our sample.

\begin{figure}
\begin{center}
\includegraphics[height=8.0cm, angle=0]{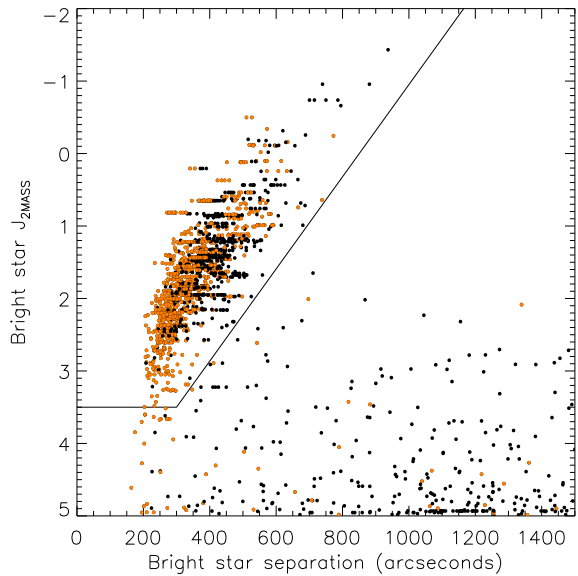}
\caption{The separation between $W2$-only sources and their nearest bright 2MASS neighbour. Over-plotted in orange are sources flagged as halo associations in the WISE All-Sky release database. The solid line delineates our own halo rejection region. \label{fig:nearby}}
\end{center}
\end{figure}

Figure \ref{fig:nearby} shows a zoomed in plot of the separations between our 3252 sources and their nearest bright 2MASS neighbour (where $J_{2MASS}$ is the 2MASS $J$-band magnitude of the bright object). The full range of separation extends out to $\sim$10,000 arcseconds, but it can be seen that a large number of candidates are found within about 600 arcseconds of a bright star if its magnitude is $J_{2MASS}\le$3.5. This over-density (1590 sources) clearly identifies the separation-brightness region where halo contamination is common amongst our $W2$-only {\tt{w2snr}}=8-10 sources. For comparison we over-plot (in orange) the 946 sources that are flagged as halo associations in the WISE database, and it is clear that the WISE database does not account for all of the $W2$-only contamination around bright stars. We therefore removed any {\tt{w2snr}}=8-10 sources that are close to a bright 2MASS star with $J_{2MASS}\le 3.5$ if their separation ($s$) from this star fulfills the criterion $s\le(-157.14\times J_{2MASS})+850$ (shown by the solid line in Figure 2). This criterion removed about half of the 3252 $W2$-only sources, which were found in only 193 square degs (0.5\%) of sky around bright stars.

\subsection{Visual inspection}
\label{sec:vis_inspect}

We used the WISE Image Service at the NASA/IPAC Infrared Science Archive to examine candidate W2-only sources that came through both our selections ({\tt{w2snr}}$>$10 and {\tt{w2snr}}=8-10). We visually examined 600 arcsecond Atlas images in all four WISE bands, but made use of the zoom and background image stretch settings in some cases. We assessed the quality of sources in several ways. We rejected sources that were associated with artefacts, including diffraction spikes, optical ghosts, glints, spurious halo associations, and some sources in regions with a poorly fit sky (which can result from latents). Next we compared the $W2$ sources with noise features in the local sky, and visually assessed their shape for point-source consistency. We thus rejected many sources that were part of resolved extended structure such as nebulosity and galaxies. We also rejected visually blended sources if the blending precluded reasonable quality assessment in other respects. And we rejected sources with visual counterparts in the $W1$, $W3$ or $W4$ bands. In some cases the presence of such counterparts was unambiguous, though in others we had to assess significance relative to noise features in the local sky.

We also used the WISE 3-Band Cryo data release to examine a second epoch of imaging data. 3-Band Cryo data was acquired by WISE following the exhaustion of solid hydrogen in the satellite's payload outer cryogen tank, and covers 30\% of the sky. Where available this data provides an additional Atlas image epoch $\sim$6 months after the All-Sky measurement. This allowed us to perform the same visual assessments on a second epoch of data, and also examine the two $W2$ epochs for any evidence of proper motion. While the spatial resolution of WISE is low (the $W2$ full-width-half-max is 6.4 arcseconds, and WISE Atlas Images have 1.375 arcseconds per pixel) nearby objects with proper motion of 2.5-3 arcseconds per year could move by $\sim$a pixel between the two epochs, and may thus be identified through visual inspection. We rejected nine sources which looked spurious in the extra epoch image.

\subsection{The new sample}
\label{sec:new_sample}

In all we have identified 158 $W2$-only sources, 52 with {\tt{w2snr}}$\ge$10 and 106 with {\tt{w2snr}}=8-10. Twenty eight of these sources are previously identified T5-Y dwarfs from the literature \citep[24 {\tt{w2snr}}$\ge$10 and four with {\tt{w2snr}}=8-10; K12;][]{burningham2010a,lodieu2012a}. Twenty four of the re-identified objects are previous WISE discoveries from the WISE census summarised by K12, so to provide a more in-depth comparison with this work we filtered our sample using the K12 selection criteria. In total 37 of our sample pass the K12 selection criteria and 121 do not. For those that do not, Table \ref{tab:comp_kp12} summarises the reasons.

\begin{table}
\begin{center}
\begin{tabular}{|c|c|c|}
\hline
               & \multicolumn{2}{|c|}{New sample}\\
               & {\tt{w2snr}}$\ge$10 & {\tt{w2snr}}=8-10 \\
\hline
Total          &     52       &      106     \\
Retained by    &     29       &      8       \\
K12 criteria   &              &              \\
Rejected by    &     23       &      98      \\
K12 criteria   &              &              \\
\hline
Why rejected?  &              &              \\
\hline
W1-W2$>$2      &      0       &      0       \\
Detections     &     21       &      98      \\
W2-W3$<$3.5    &      0       &      0       \\
Artefact       &      0       &      0       \\
Blended        &      3       &      0       \\
Profile        &      0       &      0       \\
GP             &      0       &      0       \\
$<$20pc        &      0       &      0       \\
\hline
\end{tabular}
\end{center}
\caption{Summary of how the new sample selection criteria map onto the K12 criteria. `Detections' refers to the required number of detections in individual frames. `Blended' refers to the requirement that a source is not blended with another source. `Profile' is the constraint placed on the quality of the profile-fit photometry. `GP' refers to the Galactic Plane exclusion region, and `$<$20pc' refers to the colour-magnitude constraints designed to remove objects beyond a distance of 20pc. \label{tab:comp_kp12}}
\end{table}

None of our sample are rejected by the K12 $W1-W2>2$ or $W2-W3<3.5$ criteria, unsurprising since we are considering $W2$-only sources. None of our sources are flagged as artefacts, and the K12 Galactic Plane (`GP') exclusion region covers significantly less sky (240 sq degs) than our $A_v<$0.8 criterion (8952 sq degs). Our constraints on the reduced $\chi^2$ of the $W2$ profile fit photometry are substantially tighter than those of K12 (see Section \ref{sec:rejection_methods}), so none of our sources are rejected by this K12 criterion. And the K12 `$<$20pc' criterion does not affect our sample because our sources only have lower limit $W1-W2$ colours.

A small fraction of our sources fail the K12 blending requirement. However, the K12 criterion that affects our sample the most is the `detection' criterion - K12 required that a source must have at least 8 detections ($\ge3-\sigma$) in the individual $W2$ frames, or where there are 5-7 detections it must be detected in at least 40\% of the $W2$ frames. 87\% of our {\tt{w2snr}}$>$10 sources (and all the {\tt{w2snr}}=8-10 sources) that are rejected by the K12 criteria failed to meet the K12 detection criterion.

\begin{figure*}
\begin{center}
\includegraphics[height=8.0cm, angle=0]{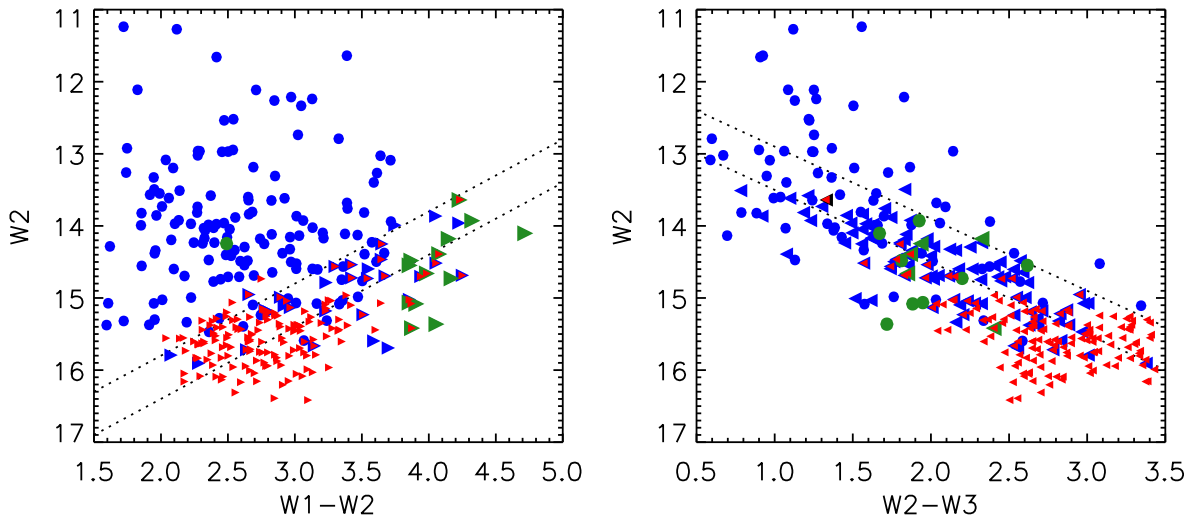}
\caption{WISE $W2$ against $W1-W2$ and $W2$ against $W2-W3$ colour magnitude diagrams. Our new sample is shown with red symbols (triangles point away from the lower limit colours). Objects from the K12 census are also shown (blue, except for Y dwarfs which are green). Dotted lines show the 5$\sigma$ and 2$\sigma$ $W1$ and $W3$ limits (assuming 20 frames). \label{fig:cmd}}
\end{center}
\end{figure*}

Figure \ref{fig:cmd} plots the new sample (red symbols) in $W2$ versus $W1-W2$ and $W2-W3$ colour-magnitude diagrams. Y dwarfs \citep[from K12 and ][]{tinney2012} are green, with other K12 sources blue. Triangles indicate lower and upper limit colours, pointing to the right and left respectively. The dotted lines indicate the 5$\sigma$ and 2$\sigma$ $W1$ and $W3$ limits (assuming 20 frames) in the left hand and right hand plots respectively. The limit colours become common in the 2-5$\sigma$ region, and dominate at fainter magnitudes.

Where our search has re-identified K12 objects the red symbols can be seen nested within the blue/green symbols. We re-identify 25 of the 33 $W2$-only detected K12 objects, with five being in $A_v>$0.8 directions, one each being rejected by our profile-fit photometry and photometric uncertainty rejection methods (see Section 2.4 and Figure 1), and one having {\tt{w2snr}}$<$8. Of the fourteen Y dwarfs one has a measured $W1$ magnitude in the All-Sky database due to an erroneous de-blending procedure as noted by K12, and eight are detected in $W3$. Most of these $W3$ detections lie around the indicated $W3$ detection limit (shown in the right hand plot), and indeed their $W3$ photometric uncertainties range from 0.2-0.5. However, three of these $W3$-detections are markedly below these lines because there are $>$30 frames of coverage for these objects (the faintest one being covered by 84 frames). Four of the five $W2$-only Y dwarfs are re-identified by our search method, with the fifth one being in an $A_v>$0.8 direction.

The brightest object in our new sample is WISE 1639-6847, a blended source with $W2$=13.64. This is the brightest (in $W2$) Y dwarf known, and was recently discovered by \citet{tinney2012} just prior to the submission of this paper. We discuss our own analysis of this object in Section \ref{sec:wise1639}. In the $W2$=14-15 range we identify 16 sources, 15 of which are previously known late T and Y dwarfs and 2 of which are rejected by the K12 selection method. The dominant magnitude range for the new sample is $W2$=15-16, where we find 116 sources. Thirteen of these are previously reported objects, with 95 being rejected by the K12 selection method. Our sample also contains 25 sources fainter than $W2$=16, 14 of which are rejected by the K12 method.

\section{Candidate classification}
\label{sec:classification}

Our candidate sample will contain many mid-late T dwarfs as well as a smaller number of Y dwarfs, and we carry out our near-infrared imaging follow-up so as to classify such objects, as well as to weed out spurious candidates. The majority of late objects in our sample will be T dwarfs, and their blue $J-H$ colour means it is faster overall to search for near-infrared counterparts in the $J$-band. To guide this process we considered the expected photometry and proper motion that our candidates should have if they are T or Y dwarfs with disk or thick-disk/halo kinematics. T5-8 dwarfs typically have $J-W2$=2-4, with T9s redder at $J-W2$=4-6 (e.g. K11). The known Y dwarfs all have $J-W2>$4 (the bluest one to date having $J-W2$=4.32$\pm$0.26) and their $J-W2$ can be extremely red (e.g. $J-W2$=9.19$\pm$0.36). Y dwarfs have M$_{W2}>$14.5 (see Fig 13 of K12), and we thus expect Y dwarfs in our sample (with W2=15-16 in our sample) to have distances $<$20pc. Mid-late T dwarfs have M$_{W2}<$14 and will thus generally be beyond $\sim$16pc (for W2=15-16). For distances beyond 16pc and with a typical disk velocity dispersion of 25-50 km~s$^{-1}$ \citep[for 1-5 Gyr, e.g. ][]{griv2009} we expect T dwarf proper motions of $<$0.7 arcsec~yr$^{-1}$. With distances $<$20pc Y dwarfs will general have proper motion $>$0.3 arcsec~yr$^{-1}$, though in some rare cases e.g. where their velocity is mostly along the line of sight, their proper motion could be smaller. Bringing this information together we defined some representative candidate classification categories to help us prioritise based on follow-up measurements of detected $J$-band counterparts. These categories are summarised in Table \ref{tab:classes}.

\begin{table*}
\begin{center}
\begin{tabular}{|l|l|l|l|}
\hline
Category & $J-W2$ & $\mu_{tot}$ & Interpretation \\
         &      & (arcsec~yr$^{-1}$) &                \\
\hline
1        & 2-4  & $<$0.7  & mid-late T (disk)              \\
2        & 2-4  & $>$0.7  & mid-late T (thick-disk/halo)   \\
3        & 4-6  & $<$0.3  & late T (disk)                  \\
4        & 4-6  & 0.3-0.7 & late T or Y (disk)             \\
5        & 4-6  & $>$0.7  & late T (thick-disk/halo) or Y  \\
6        & $>$6 & $<$0.7  & low $\mu$ Y                    \\
7        & $>$6 & $>$0.7  & clear Y                        \\
\hline
\end{tabular}
\end{center}
\caption{Candidate classification using near-infrared colour and proper motion. \label{tab:classes}}
\end{table*}

To assess the credence of possible $J$-band counterparts of our WISE candidates we considered on-sky densities of $J$-band sources. A search of VHS reveals $\sim$0.01 sources with $J$=17.5-20.2 per 7 sq arcseconds (equivalent to a circular sky area of radius 1.5 arcseconds) in directions away from the Galactic plane. It is therefore very unlikely to find mis-matched counterparts in this separation and magnitude range, and we thus assume any such counterparts to be genuine. WISE astrometric accuracy ranges from 0.28-0.44 arcseconds for $W2$-only sources with {\tt{w2snr}}=8-15 \citep[using ][]{zacharias2010,wright2010}, so sources within 1.5 arcseconds of the WISE position will have moved no more than 2 arcsec between the WISE and near-infrared epochs, and with typical baselines of a few years such sources will generally be in categories 1, 3, 4, or 6 (see Table \ref{tab:classes}).

At larger offsets mis-matched near-infrared counterparts become more likely, e.g. we find $\sim$0.6 VHS sources with $J$=17.5-20.2 per 314 sq arcseconds (equivalent to a 10 arcsecond circle of sky). We routinely check SuperCOSMOS and SDSS (where available) to assess possible offset counterparts. However, such faint sources are generally non-detections in these surveys. It is therefore important to verify widely offset counterparts through additional epoch images (showing the proper motion) or multi-band photometry giving near infrared colour criteria, before assigning them to category 2, 5 or 7. If no near-infrared counterpart is detected the candidate cannot be classified using Table \ref{tab:classes}, and additional mid-infrared follow-up will then be required.

\section{Additional photometry and proper motions}
\label{sec:phot}

We used the WFCAM Science Archive and the VISTA Science Archive (run by the Wide Field Astronomy Unit at the Institute for Astronomy, Royal Observatory Edinburgh) to search for available near-infrared imaging of our sample in Data Release 9 of the UKIDSS surveys \citep{lawrence2007}, as well as the VISTA Hemisphere Survey (McMahon and the VHS Collaboration, 2012, in preparation) and VISTA VIKING survey \citep{findlay2012}. We also searched the UKIRT Hemisphere Survey (UHS; a UKIDSS follow-on survey), taking advantage of pipeline processing and fast access to the data provided by the Cambridge Astronomical Survey Unit. These UKIRT and VISTA surveys reach J=20.2 and J=20.9 respectively. In this way we obtained near-infrared imaging for three new candidates from the UKIDSS LAS, eight new candidates from the VHS, and seven new candidates from UHS. We also obtained $J$-band imaging of four candidates and $H$-band imaging of one candidate with the Four Star instrument on the 6.5m Magellan Baade Telescope at Las Campanas Observatory, Chile. And we measured $J$-band images of seven candidates with the Near-Infrared Camera Spectrometer (NICS) instrument on the 3.58m Telescopio Nazionale Galileo (TNG) at Roque de los Muchachos Observatory on La Palma in the Canary Islands. In addition we obtained $J$- and $H$-band images of 2 candidates with the HAWK-I instrument on the European Southern Observatory (ESO) Very Large Telescope. And we measured 17 candidates using the SofI instrument on the ESO New Technology Telescope. This data provides follow-up for 45 candidates, though here we only report on the candidates we have been able to confirm (i.e. classify according to Table \ref{tab:classes}).

Four Star observations were obtained on August 10th 2012. Images were taken with the $J$-band filter (MKO system), using individual integrations of 20s, with 6 co-adds and a random 11 point jitter pattern, leading to total exposure times of 22 mins. The target was centered on detector 2 as it has the most uniform pixel sensitivity, providing a 5.4 arcmin FOV. The data were reduced using standard IRAF routines.

NICS \citep{baffa2001} observations were made on the nights of August 2nd, 3rd and 5th 2012. Image mosaics were produced by combining individual exposures of 30 seconds, with 2 or 4 co-adds, and 5 different pointings giving a total exposure time of 5 or 10 minutes. The mosaics were produced using the reduction Speedy Near-infrared data Automatic Pipeline (SNAP) provided by TNG. The TNG $J$-band filter is somewhat different to MKO $J$, so to determine a correction we derived synthetic photometry on the two systems. For standard stars \citep[from the Pickles spectral library; ][]{pickles1998} we found essentially no difference between the different $J$-band magnitudes at a level of $<$0.4\%, and could thus use the transforms of \citet{carpenter2001} to convert 2MASS $J$ into TNG $J$. For mid-late T dwarfs however, we found a significant difference. We derived synthetic photometry using spectra of mid-late T spectral standards from \citet{burgasser2006}, and determined that ($J_{TNG}-J_{MKO}$)=0.55$\pm$0.08.

HAWK-I observations were obtained on the nights of October 18th and 19th 2012 on the UT4 telescope on Paranal, Chile. HAWK-I is equipped with four Hawaii 2RG 2048x2048 pixel detectors and an MKO filter set which matches the VISTA filter set precisely \citep{kissler-patig2008}. In the $J$-band five 60s exposures were taken, with a random offset of up to 20 arcseconds between them. The $H$-band data consisted of nine 100s exposures, randomly offset from each other by up to 20 arcseconds. Data reduction was carried out using the HAWK-I pipeline (V1.8.9) provided by ESO.

SofI (Son of ISAAC) observations were obtained on 31 December 2012 under photometric conditions, seeing $\sim$0.8 arcsec, and dark conditions. SofI is equipped with a 1024x1024 Hawaii HgCdTe offering a field-of-view of 4.9$\times$4.9 arcmin and a pixel size of 0.292 arcsec in its ``Large-field'' configuration. $J$-band images were measured using integrations of 10 seconds repeated 6 times with a 10-point dither pattern yielding a total exposure of 10 min. Dome flats were obtained during the afternoon preceding the observations. The images were reduced with the ESO package {\tt{gasgano}}.

Warm-Spitzer IRAC photometry was obtained for two of the dwarfs presented in this work via Cycle 8 GO program  80077 (PI Leggett). The observations were carried out in both the [3.6] and [4.5] bands, with a 30s frame time, repeated 1 to 4 times per pointing, and dither patterns consisting of 12, 16 or 36 positions. Post-basic-calibrated-data mosaics generated by the Spitzer pipeline version S19.1.0 were used to obtain aperture photometry using a 7.2 arcseconds diameter aperture, and the sky levels were determined from annular regions. Aperture corrections were taken from the IRAC handbook, and a 3\% uncertainty (due to systematics) contributes in quadrature to overall uncertainty. The Spitzer observations and photometry are summarised in Table \ref{tab:spitzer}.

We derived proper motions using a combination of near-infrared follow-up and WISE images, using multi-epoch near-infrared images where available. When we used a WISE epoch we required near-infrared counterparts within $\sim$1.5 arcsec of the WISE position, and that the WISE source was isolated and visually un-blended to ensure reliable measurements. We typically selected 12-20 reference stars to define our inter-epoch positional transforms, which had root-mean-square residuals of $\sim$0.03-0.07 arcsecs. $J$-band counterpart centroiding uncertainties were estimated as a function of signal-to-noise, from the scatter in centroid measurements of a simulated population of sources with Gaussian point spread functions and added Poisson noise. Fit residuals and centroiding uncertainties were combined in quadrature to give full proper motion uncertainties.

To date we have been able to assign categories to eight new candidates using our J-band follow-up classification method. Six have $J$-band counterparts within 1.5 arcseconds of the WISE source centre, and two have high proper motion near-infrared counterparts. Spitzer photometry was obtained for the two high proper motion objects. The follow-up photometry, assigned categories and proper motion measurements are presented in Tables \ref{tab:phot_table}, \ref{tab:pm_table} and \ref{tab:spitzer}. Near-infrared measurements of the other followed up candidates cannot be presented at this stage because in general we have identified possible counterparts and we need an additional measurement to confirm of reject them.

\begin{table*}
\begin{center}
\begin{tabular}{|l|l|l|l|l|l|l|l|l|}
\hline
Object                    &   $W2$         & $W1-W2$ & $W2-W3$ & $J-W2$              & $H-W2$              & $J-H$                & $Y-J$               & Category \\
\hline
WISE J001354.39+063448.2  & 15.04$\pm$0.11 & $>$3.27 & $<$2.44 & 4.71$\pm$0.12       & 4.98$\pm$0.12       & -0.28$\pm$0.07       &                     & 5 \\ 
WISE J024345.58-021326.5  & 15.21$\pm$0.10 & $>$2.85 & $<$2.25 & 2.50$\pm$0.16       &                     &                      &                     & 1 \\ 
WISE J034019.34-003702.4  & 15.26$\pm$0.12 & $>$2.67 & $<$2.65 & 3.68$\pm$0.15       &                     &                      & 0.97$\pm$0.21       & 1 \\ 
WISE J041657.61-472115.0  & 16.17$\pm$0.13 & $>$2.60 & $<$2.53 & 2.41$\pm$0.14       &                     &                      &                     & 1 \\ 
WISE J083337.83+005214.2  & 14.96$\pm$0.10 & $>$3.39 & $<$2.40 & 5.32$\pm$0.14       & 5.67$\pm$0.14       & -0.35$\pm$0.14       & 0.15$\pm$0.24       & 5 \\ 
WISE J132116.09-100512.3  & 15.72$\pm$0.14 & $>$3.00 & $<$2.50 & 3.73$\pm$0.19       &                     &                      & 1.34$\pm$0.33       & 1 or 2$^a$ \\ 
WISE J140143.50+394235.2  & 15.44$\pm$0.10 & $>$3.33 & $<$2.22 & 2.90$\pm$0.16       &                     &                      &                     & 1 \\ 
WISE J210216.60-272828.8  & 15.36$\pm$0.14 & $>$2.88 & $<$2.80 & 2.40$\pm$0.15       & 2.85$\pm$0.18       & -0.45$\pm$0.13       & 1.53$\pm$0.11       & 1 or 2$^a$ \\ 
\hline
\multicolumn{9}{|l|}{$^a$Ambiguous due to lack of proper motion constraint.}\\
\end{tabular}
\end{center}
\caption{Photometry of new late object discoveries. Mid-infrared limit colours are also given where one band has only a 95\% confidence upper limit magnitude. The object category is listed in the last column based on colour and proper motion (see Table \ref{tab:classes}).\label{tab:phot_table}}
\end{table*}

\begin{table*}
\begin{center}
\begin{tabular}{|l|l|l|l|l|l|l|}
\hline
Object         & Facility       & Band & Obs-date   & $\mu_{\alpha cos\delta}$ & $\mu_{\delta}$ & RPM$_{W2}^a$ \\
               & (epoch 1 \& 2) &      &            & (arcsec~yr$^{-1}$) & (arcsec~yr$^{-1}$) &      \\
\hline
WISE 0013+0634 & UKIDSS LAS     & $J$  & 2008/11/15 & +1.17$\pm$0.09 & -0.54$\pm$0.09 & 20.60        \\ 
               & Magellan       & $J$  & 2012/08/10 &                &                &              \\
WISE 0243-0213 & WISE-All-Sky   & $W2$ & 2010/01/25 & +0.21$\pm$0.41 & -0.03$\pm$0.39 & -            \\ 
               & TNG            & $J$  & 2012/08/03 &                &                &              \\
WISE 0340-0037 & UKIDSS LAS     & $J$  & 2005/10/26 & +0.00$\pm$0.06 & -0.03$\pm$0.06 & -            \\ 
               & VISTA VHS      & $J$  & 2011/01/03 &                &                &              \\
WISE 0416-4721 & WISE-All-Sky   & $W2$ & 2010/01/29 & -0.27$\pm$0.03 & +0.19$\pm$0.03 & 18.76        \\ 
               & NTT            & $J$  & 2012/12/31 &                &                &              \\
WISE 0833+0052 & UKIDSS LAS     & $J$  & 2006/12/11 & +0.86$\pm$0.04 & -1.68$\pm$0.04 & 21.34        \\ 
               & VLT            & $J$  & 2012/10/18 &                &                &              \\
WISE 1401+3942 & WISE-All-Sky   & $W2$ & 2010/06/26 & -0.27$\pm$0.36 & 0.08$\pm$0.37  & -            \\ 
               & TNG            & $J$  & 2012/08/06 &                &                &              \\
WISE 1639-6847 & WISE-All-Sky   & $W2$ & 2010/03/10 & -0.8$\pm$1.2$^b$ & -2.8$\pm$1.2$^b$ & 20.80     \\ 
               & WISE-3BC       & $W2$ & 2010/09/06 &                &                &              \\
\hline
\multicolumn{7}{|l|}{$^a$ RPM is presented where proper motion is inconsistent with zero.}\\
\multicolumn{7}{|l|}{$^b$ See \citet{tinney2012} for more accurate proper motions.}\\
\end{tabular}
\end{center}
\caption{Proper motion and reduced proper motion for late type objects discussed in this paper.\label{tab:pm_table}}
\end{table*}

\begin{table*}
\begin{center}
\begin{tabular}{|l|l|l|l|l|l|}
\hline
Object               & Date       & Integration & [3.6]           & [4.5]         & [3.6]-[4.5]   \\
                     & observed   & time (s)    &                 &               &               \\
\hline
WISE 0013+0634       & 2013-02-12 & 1440        & 17.15$\pm$0.03 & 15.16$\pm$0.03 & 1.99$\pm$0.04 \\
WISE 0833+0052       & 2013-01-03 & 1440        & 17.02$\pm$0.03 & 14.80$\pm$0.03 & 2.22$\pm$0.04 \\
\hline
\end{tabular}
\end{center}
\caption{Spitzer IRAC Photometry of new late object discoveries.\label{tab:spitzer}}
\end{table*}

\section{Spectroscopy}
\label{sec:spectra}
Low-resolution near-infrared spectroscopy of WISE J001354.39+063448.2 (WISE 0013+0634 hereafter) and WISE J083337.83+005214.2 (WISE 0833+0052 hereafter) were obtained using the Gemini Near Infrared Spectrograph \citep[GNIRS; ][]{elias2006} on the Gemini North Telescope on Mauna Kea, Hawaii, and the Folded port InfraRed Echellette (FIRE) spectrograph \citep{fire08,fire10} mounted on the Baade 6.5m Magellan telescope at Las Campanas Observatory respectively. The FIRE and GNIRS observations were made on the nights of 14$^{th}$ February and 25 June 2013 in clear conditions.

GNIRS observations were made via queue program GN-2013B-Q-65. A 0.68 arcsecond slit was used in cross-dispersed mode providing a 0.9 to 2.5 micron spectrum with  a resolving power of R$\sim$750. Eight 300 second integrations were obtained using an ABBA jitter pattern giving a total integration of 40 minutes. Comparison argon arc frames were obtained to provide dispersion solutions. For telluric correction the F5V star HD 5892 was observed after the target and at a similar airmass. Data reduction was carried out using standard IRAF Gemini packages \citep{cooke2005}.  Hydrogen lines were removed from the  F5V stellar spectrum using linear interpolation, and the true continuum determined using the IRTF Spex library spectrum of HD 27524 \citep{Rayner2009}.

The FIRE instrument was used in prism mode with a 0.6\arcsec slit giving a resolving power of R$\sim 200$ at 1.6$\micron$. Thirty 120~second integrations were obtained using an ABBA jitter pattern, giving a total integration of one hour. Ne-Ar arc lamps were observed at the configuration and pointing for both the target and the standard to allow for good dispersion correction. Due to some saturation with the on-night standard calibration star we made use of calibrators from 5$^{th}$ October and 3$^{rd}$ June 2012. These nights had similar conditions to that of our target observation, and the standards were observed at a similar airmass. To verify the effectiveness of this approach we generated four output spectra of WISE0833+0052 using the four measured standards in order to compare the results and assess consistency. We found $\le$10\% variation in flux calibration over the full wavelength range using the different standard stars, and significantly less variation over the wavelength width of the flux peaks that we measure in Section \ref{sec:wise0013_0833}.

The FIRE spectrum was extracted using the low-dispersion version of the FIREHOSE pipeline, which is based on the MASE pipeline \citep{mase,fire10}. The pipeline uses a flat field constructed from two quartz lamp images taken with the lamp at high (2.5 V) and low (1.5 V) voltage settings. The data were divided by this pixel flat before being wavelength calibrated. The pipeline performs sky subtraction following the method outlined in \citet{bochanski2011}, adapted for the low-dispersion configuration of the spectrograph. The spectra were optimally extracted before being combined using a weighted mean, using an adaptation of the xcombspec routine from SpexTool \citep{cushing2004}. The T dwarf spectra were then corrected for telluric absorption and flux calibrated using a FIRE specific version of the xtellcor routine \citep{vacca2003}. Finally, residual outlying points due to cosmic rays and bad pixels were removed using a simple 3-sigma clipping algorithm.

\section{Objects of special note}
\label{sec:new}

\subsection{WISE 1639-6847}
\label{sec:wise1639}

The brightest (in $W2$) source in our sample is WISE 1639-6847, which we made a study of while assessing sample candidates. This object was recently presented by \citet{tinney2012}, who used differential imaging to separate the $J$-band counterpart from a nearby blended star $\sim$6 arcsec to the south of the object (measuring a colour of $J_3-W2$=6.98). These authors also measured the object's spectrum during a period of good ($\sim$0.6 arcsec) seeing, and determined a spectral type of Y0. They also combined the WISE All-Sky and 3-Band Cryo images with their $J$-band data to measure proper motion and parallax constraints, determining a proper motion of 3.07$\pm$0.04 arcsec~yr$^{-1}$ and an approximate distance of $\sim$5pc. We did not measure the near-infrared counterpart, but we used the WISE data by itself to assess source motion and determine proper motion constraints. It is interesting to compare our results to those obtained with more data.

Examination of the blended source in 2MASS, SuperCOSMOS and $W1$ revealed that it was a non-moving point source with $\sim$K-star colours. To remove the blended source from the $W2$ image we used PSF-subtraction. We identified an isolated PSF-subtraction source $\sim$30 arcsecs south-west of the blended source, extracted a sky-subtracted $W2$ image region centred on PSF-subtraction source (and absent of any other sources), normalised this to account for the $W2$ flux difference between the blended source and the PSF-subtraction source, and made a spatial shift to the blended star WISE position (which is well measured from the $W1$ image). This normalised shifted PSF-subtraction source was then subtracted off the $W2$ image to effectively remove the PSF of the blended source.

\begin{figure}
\begin{center}
\includegraphics[width=8.0cm, angle=0]{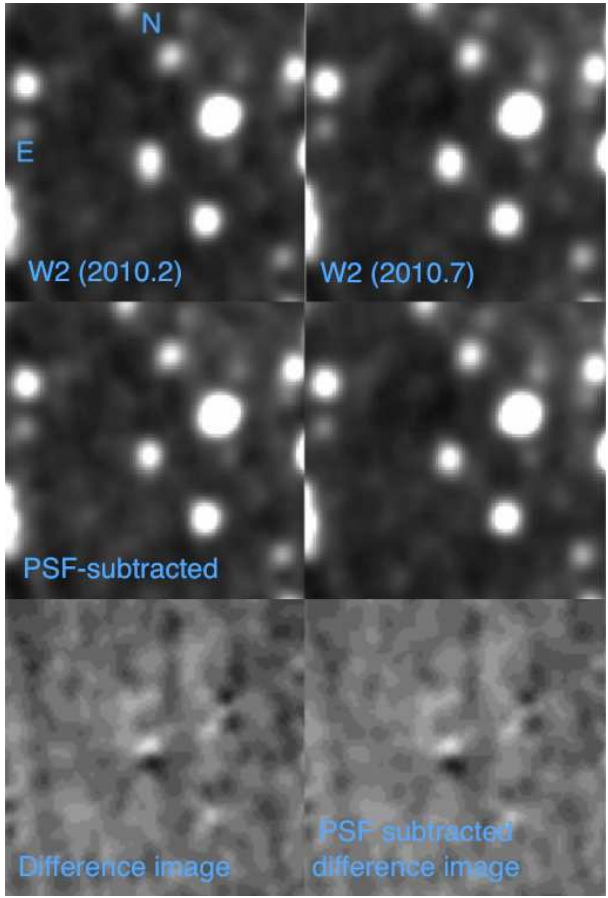}
\caption{$W2$ images of WISE 1639-6847. The top two images show the WISE All-Sky and 3-Band Cryo images on the left and right respectively, which are separated in epoch by $\sim$6 months. The PSF of the non-moving blend star has been subtracted in the middle two images, with WISE 1639-6847 now seen to be a point source that moves $\sim$1 pixel southwards between these epochs. This motion is also shown by the difference images in the bottom two plots (see text). \label{fig:mag09}}
\end{center}
\end{figure}

In Figure \ref{fig:mag09} the top left and right $W2$ images are from the WISE-All-Sky and 3-Band Cryo releases respectively ($\sim$6 months apart). The middle two plots show our blend-subtracted images where WISE 1639-6847 now appears as a circular point-like source. The proper motion was measured using these blend subtracted images. We used the IRAF, GEOMAP and GEOXYTRAN packages to derive positional transforms between the two epoch images. As reference stars we selected sources that had good signal-to-noise ($>$20), were unsaturated and point-like in both epoch images, and were within 5 arcmins of the candidate. We selected 20 reference stars, optimized a fit geometry of second order polynomials in $x$ and $y$, and rejected outlier reference stars. The root-mean-square residuals of our fitted transform were 0.05 pixels ($\sim$70mas). To estimate centroiding uncertainties for WISE sources we used the astrometric relation described in \citet{wright2010} with a correction factor of FWHM$_{W2}$/FWHM$_{W1}$ since our sources are $W2$-only ($\sigma = 0.18 + 6.4/(2 SNR)$). In this way we determined proper motion constraints for WISE 1639-6847 of $\mu_{\alpha}$=-0.8$\pm$1.2 arcsec~yr$^{-1}$ and $\mu_{\delta}$=2.8$\pm$1.2 arcsec~yr$^{-1}$. These have much larger uncertainties than the Tinney et al. results but are consistent, and significant at the level of $\sim$2.5$\sigma$.

We also studied the signature of motion in the WISE data by assessing the difference image (epoch 1 - epoch 2) which is shown at the bottom of Figure \ref{fig:mag09}. The left hand image shows the difference between the original WISE epoch images while the other image was made using the blend-subtracted versions. These two images are very similar as expected, since the subtraction of the stationary blend star should not significantly affect the difference image. The motion of the source between the epochs results in a positive-negative undulation following the direction of motion. We note other features in the difference images associated with bright sources, whose profiles are not stable between the epochs, and also some that result from faint non-point-like features. However, the difference signature of WISE 1639-6847 is clear, and it seems likely that the depth and symmetry properties of such difference image undulations could be combined with source brightness information to provide a general means of assessing the credence of moving object candidates in WISE multi-epoch data.

\subsection{WISE 0013+0634 and WISE 0833+0052}
\label{sec:wise0013_0833}

\begin{figure}
\begin{center}
\includegraphics[width=8.0cm, angle=0]{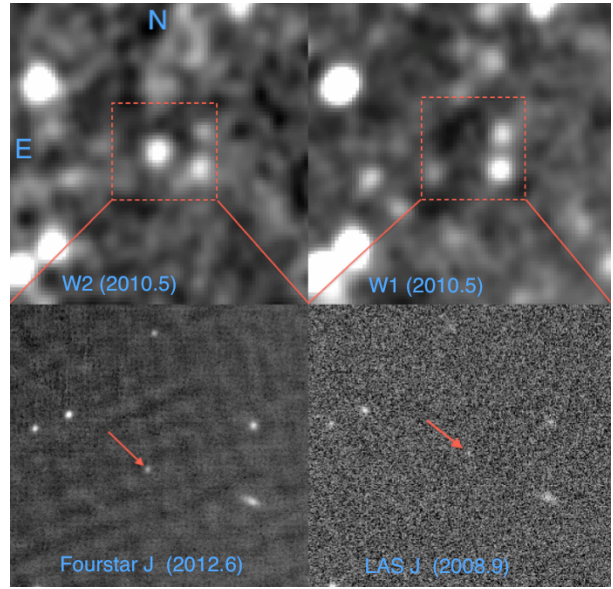}
\caption{Multi-band images of WISE 0013+0634. The top two plots are WISE $W2$- and $W1$-band images on the left and right respectively, with the source in the centre. The bottom two plots show the Magellan Four Star and UKIDSS LAS $J$-band zoomed in images on the left and right respectively. The epoch difference between these two images is 3.7 years. The high proper motion $J$-band counterpart is indicated in each image by a red arrow. \label{fig:mag01}}
\end{center}
\end{figure}

Figure \ref{fig:mag01} shows images of WISE 0013+0634 in various bands. The $W2$ and $W1$ WISE images are shown upper left and right respectively (epoch 2010.5), with two epochs of $J-$band imaging below with the near infrared counterpart indicated by a red arrow. The bottom right zoomed in image shows a low signal-to-noise ($\sim$5) counterpart from the UKIDSS LAS (epoch 2008.9). This counterpart is offset by $\sim$2 arcseconds from the WISE source centre, and no detections were discernible in the $Y$-, $H$- or $K$-band LAS images. We were able to confirm the near-infrared counterpart in a deeper $J$-band image (using Four Star on Magellan; epoch 2012.6) in which we detected the counterpart with signal-to-noise$\sim$20 on the other side of the WISE source centre. The three epoch positions (UKIDSS, WISE and Magellan) are all consistent with our measured proper motion (to within astrometric uncertainties), though our final proper motion of 1.3 arcsec~yr$^{-1}$ was derived using the UKIDSS and Magellan data (with a 3.7 yr baseline).

\begin{figure}
\begin{center}
\includegraphics[width=8.0cm, angle=0]{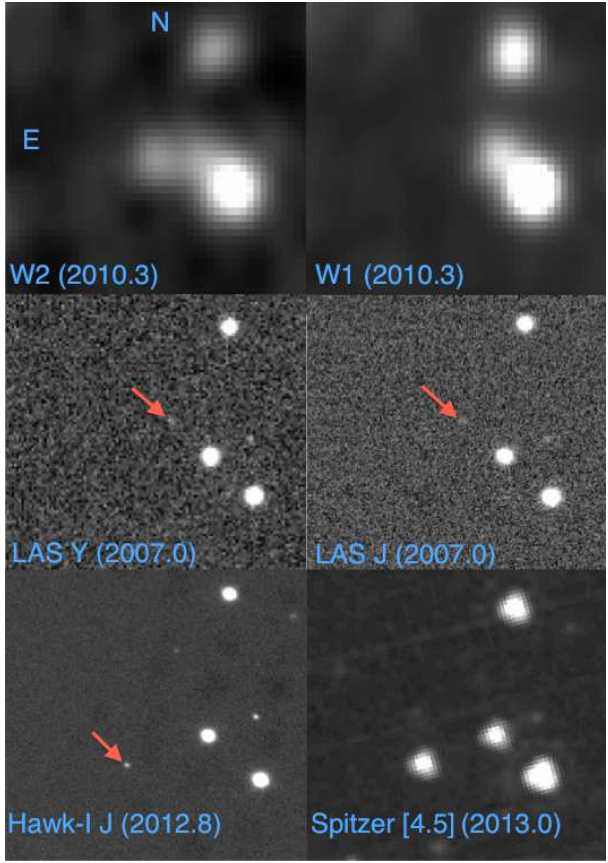}
\caption{Multi-band images of WISE 0833+0052. Each image is 1 arcminute on the side. The top two plots are WISE $W2$- and $W1$-band images on the left and right respectively, with WISE 0833+0052 in the centre. The middle two plots show the UKIDSS $Y$ and $J$-band images from 2006.95. The bottom two plots show the VLT Hawk-I $J$ image from 2012.8 and the Spitzer [4.5] image from 2013.0. The high proper motion near-infrared counterpart is indicated in each image by a red arrow, and has moved by 10.4 arcseconds across the near-infrared epochs. \label{fig:0833}}
\end{center}
\end{figure}

WISE 0833+0052 is the brightest ($W2$=14.96) new object in our sample. Figure \ref{fig:0833} shows the WISE images at the top (epoch 2010.32) with the WISE counterpart slightly blended with a star $\sim$10 arcsec to the west. The middle two plots show the UKIDSS LAS $Y$- and $J$-band images from 2007.0 (the counterpart was un-detected in the LAS $H$- and $K$-band images), and the bottom two plots show the $J$-band Hawk-I image from 2012.8 and the Spitzer [4.5] image from 2013.0. A faint UKIDSS $Y$- and $J$-band counterpart is visible at low signal-to-noise about 6 arcsecs to the north west of the WISE source centre. And in the more recent VLT and Spitzer epochs the source is strongly detected about 4 arcsecs to the south-east of the WISE position. We measured the proper motion of WISE 0833+0052 using the UKIDSS and HAWK-I $J$-band images. We also verified that the positions in UKIDSS, WISE and Spitzer were consistent with this. Our centroiding on the WISE source was not affected by the modest level of blending evident in the $W2$ image, so our consistency check was able to verify the relative location of the object in all the epochs. WISE 0833+0052 has a high proper motion of 1.89$\pm$0.04 arcsec~yr$^{-1}$.

\begin{figure}
\begin{center}
\vspace{-7.5cm}
\includegraphics[width=15.0cm, angle=0]{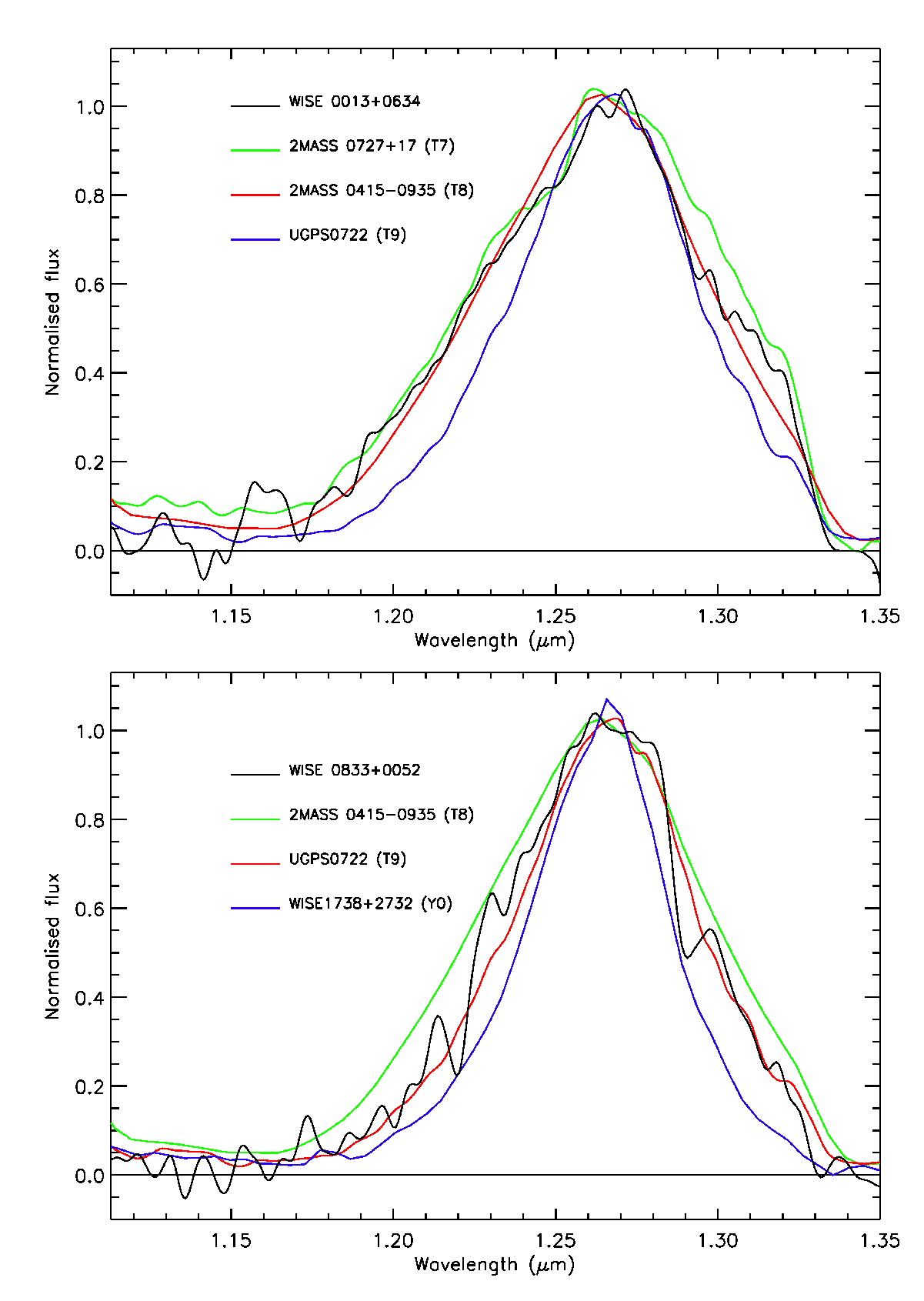}
\caption{The J-band spectra of WISE 0013+0634 and WISE 0833+0052. Late T standard spectra are over-plotted for comparison and to facilitate spectral typing.\label{fig:0013_0833_jspec}}
\end{center}
\end{figure}

Figure \ref{fig:0013_0833_jspec} shows the $J-$band spectra of WISE 0013+0634 and WISE 0833+0052. In the top plot we compare to the spectra of three normal T dwarfs with known spectral types; T7 (2MASS 0727+1710), T8 (2MASS 0415-0935) and T9 (UGPS 0722-0540). Similarly, in the bottom plot we compare with WISE J1738+2732 (Y0) in addition to the T8 and T9 examples. All spectra have been normalised to an average of unity in the 1.265-1.275 micron spectral range. The $J-$band flux peak is sculpted by strong CH$_4$ and H$_2$O absorption, and its width is a well established diagnostic for low $T_{\rm eff}$ amongst late L, T and early Y dwarfs (e.g. Burgasser et al 2006; Burningham et al. 2008; Kirkpatrick et al. 2012). We focus on the $J-$band peaks for spectral typing since this region provides the best signal-to-noise in our spectra. In the top plot it can be seen that the $J-$band flux peak of WISE 0013+0634 is most similar to the T8 comparison. The $J-$band flux peak of the T9 comparison has a significantly steeper blue wing, and the red wing of the T7 comparison flux peak is significantly brighter in the 1.28-1.32 micron range. We thus assign a spectral type of T8$\pm$0.5 to WISE 0013+0634. In the bottom plot the $J-$band flux peak of WISE 0833+0052 has a very similar width and shape to that of UGPS 0722-0540, and is clearly intermediate between (and significantly different to) the T8 and Y0 comparison spectra. We thus assign a spectral type of T9$\pm$0.5 to WISE 0833+0052.

\begin{figure}
\begin{center}
\vspace{-7.5cm}
\includegraphics[width=15.0cm, angle=0]{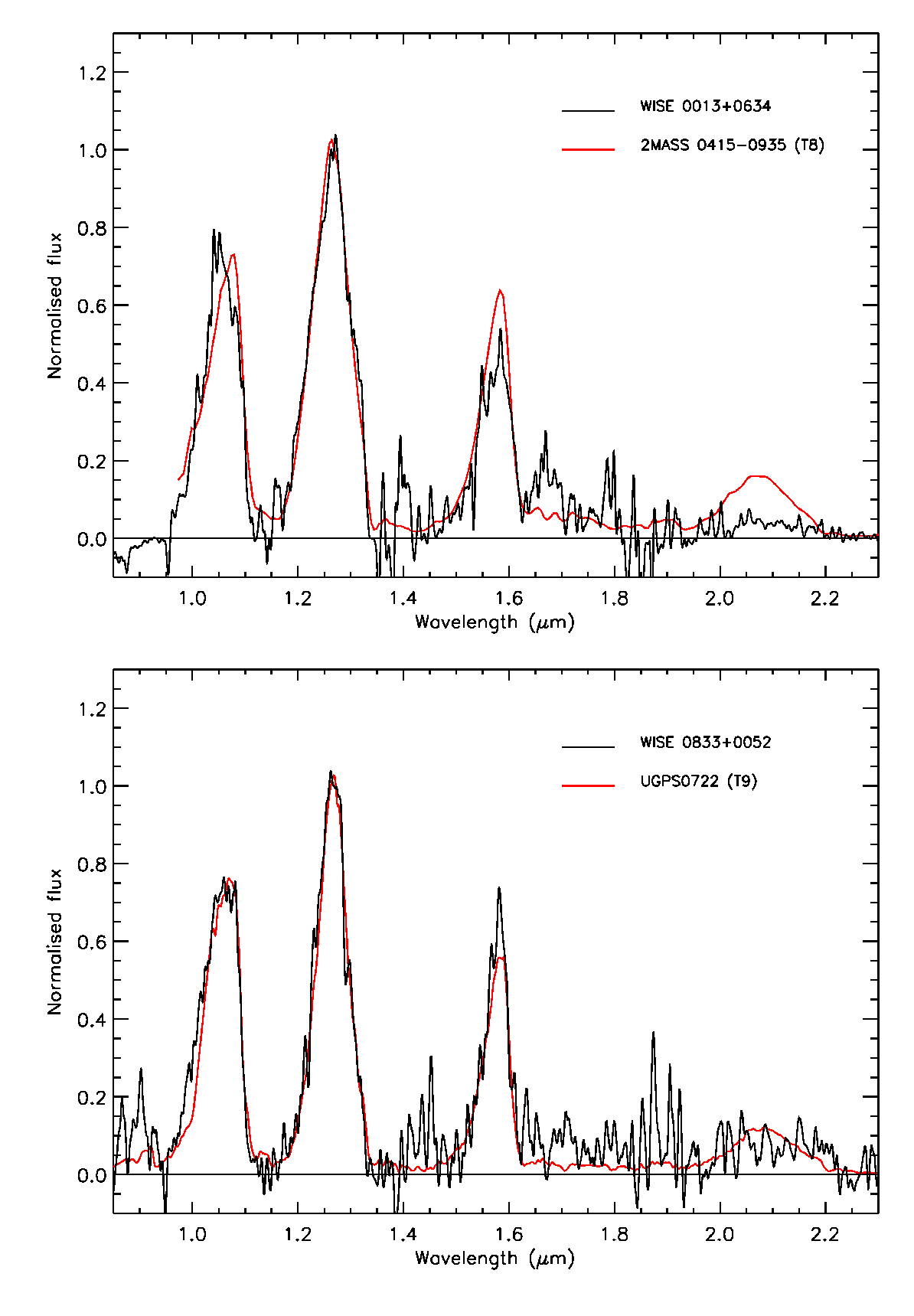}
\caption{The full near-infrared spectrum of WISE 0013+0634 and WISE 0833+0052. A known T8 and T9 dwarf are over-plotted in red for comparison. \label{fig:0013_0833_fullspec}}
\end{center}
\end{figure}

\begin{figure*}
\begin{center}
\includegraphics[width=15.0cm, angle=0]{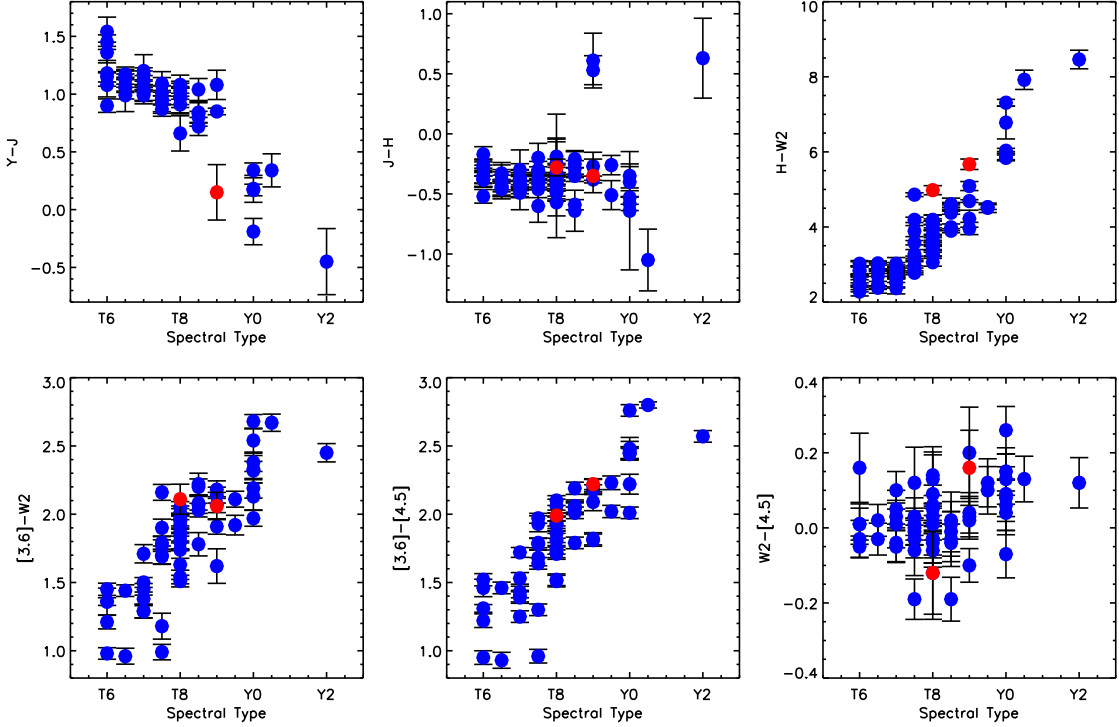}
\caption{Plots of various colours as a function of spectral type. WISE 0013+0634 (T8) and WISE 0833+0052 (T9) are red, and the comparison sample of late T and Y dwarfs from Leggett et al. 2013 is in blue. \label{fig:0013_0833_colours}}
\end{center}
\end{figure*}

To compare the overall spectral morphology and colours of WISE 0013+0634 and WISE 0833+0052 with those of other field objects, we show their full near-infrared spectra and measured colours in Figures \ref{fig:0013_0833_fullspec} and \ref{fig:0013_0833_colours}. When plotting the spectra we also show (in red) comparison T8 and T9 spectra for WISE 0013+0634 and WISE 0833+0052 respectively. In the colour plots WISE 0013+0634 and WISE 0833+0052 are red symbols, with a comparison sample of late T and Y dwarfs from \citet{leggett2013} plotted in blue.

We note that the $Y-$band flux peak of WISE 0833+0052 is not markedly different from that of UGPS 0722-0540, which seems surprising since the measured $Y-J$ colour of WISE 0833+0052 is bluer than all the late T dwarfs in Figure \ref{fig:0013_0833_colours}. While the $Y-$band peak of WISE 0833+0052 is slightly broader (extending to the blue), this can only account for a small $\sim$7\% $Y-$band excess (relative to $J$), which is insufficient to explain the measured $Y-J$ colour. However, the $Y-$band photometric uncertainties are large due to low signal-to-noise ($\pm$0.2), and the measured $Y-J$ colour is only $\sim$2-$\sigma$ away from a more typical value for late T dwarfs. So while this potential anomaly may offer tentative evidence of variability, further photometric studies (at higher signal-to-noise) will be needed to investigate this possibility.

WISE 0013+0634 is significantly $K-$band suppressed by comparison with 2MASS 0415-0935, and despite a rather noisier $K-$band spectrum WISE 0833+0052 appears to be $K-$band suppressed also. Theory and observation suggest this is due to strong collisionally induced absorption by H2 in higher pressure atmospheres with low-metallicity and/or high-gravity \citep[e.g. ][]{saumon1994,pinfield2008}.

The $H-$band flux peak of WISE 0013+0634 appears slightly suppressed relative to 2MASS 0415-0935, and that of WISE 0833+0052 seems slightly enhanced compared to UGPS 0722-0540. However, there is no evidence that they are outliers in $J-H$ colour, and we note that the broad-band spectroscopic flux calibration for WISE 0833+0052 is no better than $\sim$10\%.

The range of plots shown in Figure \ref{fig:0013_0833_colours} clearly demonstrate that WISE 0013+0634 and WISE 0833+0052 have relatively enhanced mid-infrared emission compared to their near-infrared brightness. This excess is seen in the $H-W2$ versus spectral type plot, where both objects lie significantly above the locus of other T dwarfs ($\sim$1-1.5 magnitudes redder than the average T dwarf colour). However, in $J-H$ and mid-infrared colours they do not appear unusual. The only comparison object with a similar excess is the T7.5p dwarf ULAS 1416+13B, which is believed to be a low metallicity/high gravity object \citep{burningham2010}. Theory \citep{allard2011,saumon2008} predicts a change in $H-W2$ of +0.2 to +0.3 for a decrease in metallicity of $\Delta$[M/H]=-0.3 (at $T_{\rm eff}\sim$700~K), while studies of metal poor benchmarks \citep[e.g. ][]{pinfield2012} suggest even larger effects \citep[$\Delta H-W2$=+0.6; ][]{burningham2013}. Overall, current theory and observation would suggest that $\Delta H-W2$=1-1.5 could suggest a low metallicity object with [M/H] between -0.5 and -1.5.

We have estimated the distances of WISE 0013+0634 and WISE 0833+0052 using absolute magnitude against spectral type relations from \citet{dupuy2012}, in all available bands (and allowing for a spectral type uncertainty of $\pm$0.5 types). The results are shown in Table \ref{tab:0013_0833dist} for both the single object case and that of an unresolved binary with equal brightness components. For both objects these distance constraints fall into two groupings. Distance estimates made in the $J$- and $H$-bands are the largest, with those made in the mid-infrared (and the $Y-$band for WISE 0833+0052) being smaller by a factor of $\sim$2. As previously discussed, the $Y-$band magnitude has large uncertainty, but the factor of $\sim$2 difference between the $JH$ and mid-infrared distance estimates is clear, and in itself is representative of a $\sim$1.5 magnitude relative mid-infrared/$JH$ excess (compared to typical T8-9 field dwarfs). The actual distances of WISE 0013+0634 and WISE 0833+0052 may be intermediate between the near- and mid-infrared estimates, and will be greater if the object is an unresolved multiple.

\begin{table}
\begin{center}
\begin{tabular}{|l|c|c|}
\hline
WISE 0013+0634 & \multicolumn{2}{|c|}{Distance range (pc)}\\
(T8$\pm$0.5)   &               &                          \\
Band           & Single object & Unresolved binary        \\
\hline
J              & 46 (39-55)    & 65 (55-78)               \\
H              & 43 (36-51)    & 61 (51-73)               \\
{[}3.6]        & 26 (23-30)    & 37 (33-42)               \\
{[}4.5]        & 22 (20-25)    & 31 (28-35)               \\
W2             & 21 (18-25)    & 30 (25-35)               \\
\hline
WISE 0833+0052 & \multicolumn{2}{|c|}{Distance range (pc)}\\
(T9$\pm$0.5)   &               &                          \\
Band           & Single object & Unresolved binary        \\
\hline
Y              & 22 (18-26)    & 31 (26-37)               \\
J              & 31 (26-37)    & 44 (37-52)               \\
H              & 32 (27-38)    & 45 (37-53)               \\
{[}3.6]        & 17 (15-19)    & 24 (21-27)               \\
{[}4.5]        & 15 (14-17)    & 22 (20-24)               \\
W2             & 17 (15-20)    & 24 (21-29)               \\
\hline
\end{tabular}
\end{center}
\caption{Distance estimates for WISE 0013+0634 and WISE 0833+0052 using absolute magnitude versus spectral type relations in the Y, J, H, [3.6], [4.5] and W2 bands (where available), and spectral types of T8$\pm$0.5 and T9$\pm$0.5 respectively. Absolute magnitudes have been estimated assuming single objects and also equal brightness unresolved binarity. \label{tab:0013_0833dist}}
\end{table}

\begin{figure*}
\begin{center}
\includegraphics[height=16.0cm, angle=0]{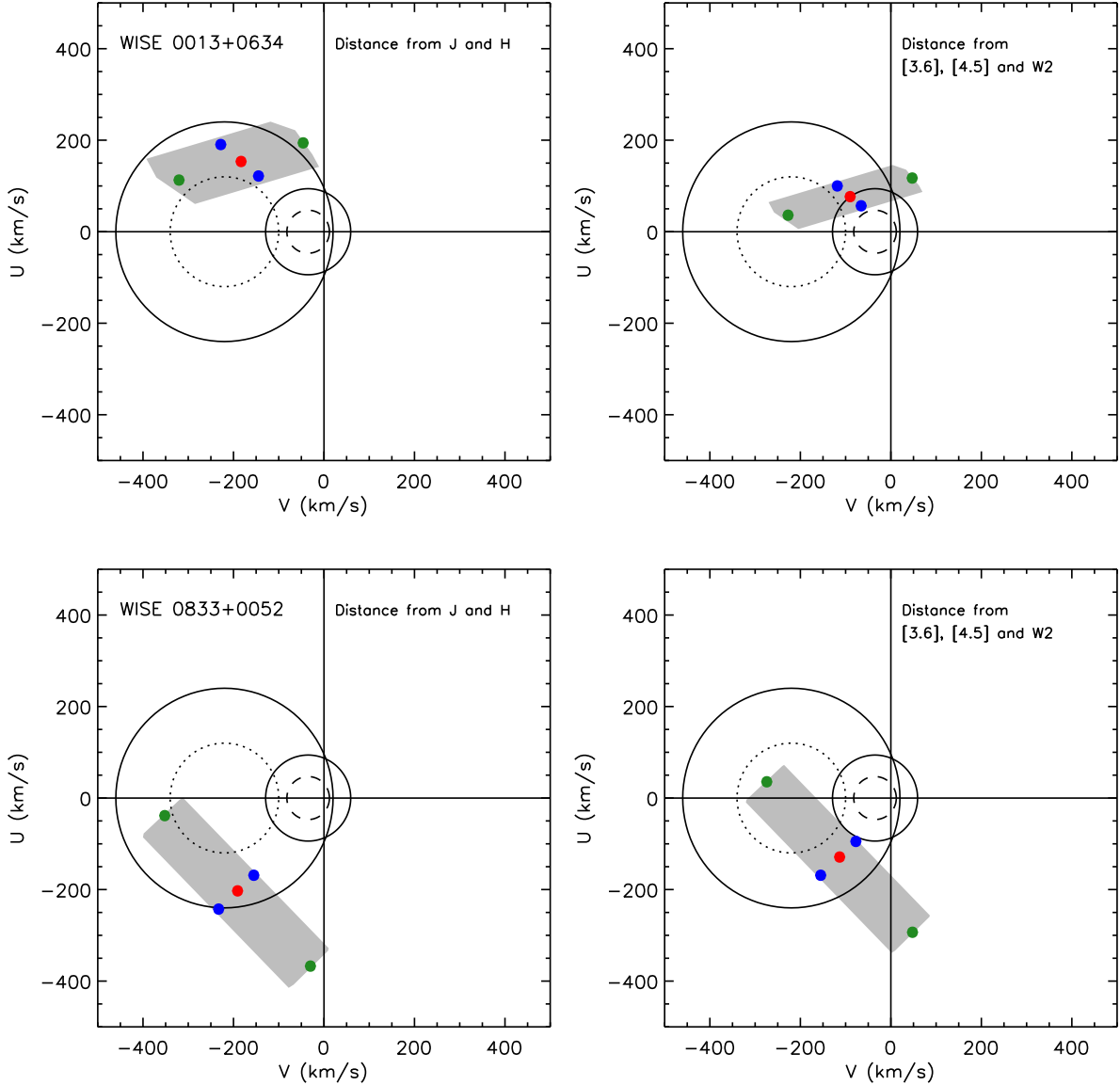}
\caption{UV space motion plots for WISE 0013+0634 and WISE 0833+0052. Uncertainties in measured proper motion and estimated distance have been accounted for, as well as a possible range in radial velocity from -250 to +250km~s$^{-1}$. In the left hand plots distances were estimated using $JH$ brightness, while in the right hand plots they were estimated using (relatively over-luminous) mid-infrared brightness. Old disk and halo velocity dispersions (1 and 2 $\sigma$ as dashed/dotted and solid lines respectively) are also shown from \citet{chiba2000}. \label{fig:0013_0833_spacemotion}}
\end{center}
\end{figure*}

We now use our distance estimates to place constraints on the space motions of WISE 0013+0634 and WISE 0833+0052. We use the single object distances to avoid over-estimating space motion, and we combine together our distance estimates into a $JH$ constraint (D$_{JH}$) and a mid-infrared constraint (D$_{MIR}$). WISE 0013+0634 thus has D$_{JH}$=36-55pc and D$_{MIR}$=18-30pc, while WISE 0833+0052 has D$_{JH}$=27-37pc and D$_{MIR}$=15-19pc. We have used these distance estimates to construct the space motion diagrams shown in Figure \ref{fig:0013_0833_spacemotion}. The velocity components $U$, $V$, and $W$ are directed to the Galactic anti-center, rotation direction, and north Galactic pole respectively, and have been corrected for the local solar motion ($U_{\odot}$, $V_{\odot}$, $W_{\odot}$) = (-8.5, 13.38, 6.49) km~s$^{-1}$ with respect to the local standard of rest \citep{coskunoglu2011}. WISE 0013+0634 is shown in the top plots, with WISE 0833+0052 in the bottom plots. The left hand plots were constructed using the $JH$ distances, while the right hand plots use the closer mid-infrared estimates. We also allow for proper motion uncertainties (which are of relatively minor importance) and possible radial velocity in the range -250~km$s^{-1}$ to +250~km$s^{-1}$ when calculating space motions, with the resulting constraints being shown as grey regions in the plots. As a guide we over-plot the mid-value space motion in red, with blue and green symbols indicating how the uncertainty in distance and radial velocity broadens the constrained region. We also over-plot the old disk and halo velocity dispersions (1 and 2 $\sigma$ as dashed/dotted and solid lines respectively) from \citet{chiba2000}. In Table \ref{tab:0013_0833space} we present tangential velocity estimates for WISE 0013+0634 and WISE 0833+0052 (with ranges reflecting proper motion and distance uncertainties), as well as minimum possible value estimates for total space motion ($\sqrt{(U^2+V^2+W^2)}_{min}$ in the LSR frame). These minimum space motions result from a $UVW$ parameter-space search (i.e within the grey regions in Figure \ref{fig:0013_0833_spacemotion}) for each object. The solar-centric radial velocities (RV) that would produce the minimum space motions are indicated in brackets in the table, with all other RV values leading to larger space motions.

\begin{table*}
\begin{center}
\begin{tabular}{|l|c|c|c|c|}
\hline
Object         & \multicolumn{2}{|c|}{V$_{tan}$ (km~s$^{-1}$)} & \multicolumn{2}{|c|}{Minimum space motion (km~s$^{-1}$)} \\
               &                 &                             & \multicolumn{2}{|c|}{=$\sqrt{(U^2+V^2+W^2)}_{min}$}\\
\hline
               & Using D$_{JH}$  & Using D$_{MIR}$             & Using D$_{JH}$ & Using D$_{MIR}$ \\
\hline
WISE 0013+0634 & 278 (213-349)   & 147 (107-190)               & 183 (RV=-1) &  83 (RV=-1) \\
WISE 0833+0052 & 286 (235-337)   & 179 (126-231)               & 222 (RV=12) & 118 (RV=12) \\
\hline
\end{tabular}
\end{center}
\caption{Tangential velocities and minimum values of the total space motion ($\sqrt{(U^2+V^2+W^2)}_{min}$) for WISE 0013+0634 and WISE 0833+0052. The solar-centric radial velocity (RV in km~s$^{-1}$) required to produce the minimum space motion is also listed. \label{tab:0013_0833space}}
\end{table*}

It is clear that when the $JH$ distances are used, WISE 0013+0634 and WISE 0833+0052 occupy the halo region in Figure \ref{fig:0013_0833_spacemotion}. This is also evident in Table \ref{tab:0013_0833space}, where the minimum possible space motions (if we assume D$_{JH}$) are above the range defined for the thick disk \citep[85-180 km~s$^{-1}$; ][]{fuhrmann2000,feltzing2003,nissen2004}. However, if we instead use the mid-infrared distance estimates the space motions could be lower, touching or overlapping with the 2-$\sigma$ old disk velocity dispersion in Figure \ref{fig:0013_0833_spacemotion}. In this case the total space motion would also be consistent with thick disk membership.

The actual distances of WISE 0013+0634 and WISE 0833+0052 may be intermediate between D$_{JH}$ and D$_{MIR}$ \citep[e.g. consider the measured absolute magnitudes of the low metallicity/high gravity dwarf ULAS~1416+13B; ][]{dupuy2012}, with space motions that are similarly intermediate. Taking an average (of the D$_{JH}$ and D$_{MIR}$ cases) minimum space motion for WISE 0013+0634 gives 133 km~s$^{-1}$, while for WISE 0833+0052 the average is 170 km~s$^{-1}$. Parallax distance and radial velocity will be crucial measurements for both these objects to properly establish their space motions, but it seems likely even at this stage that WISE 0833+0052 could have unambiguous halo kinematics.

\section{Reduced proper motion diagram}
\label{sec:rpmd}

\begin{figure*}
\begin{center}
\includegraphics[height=16.0cm, angle=0]{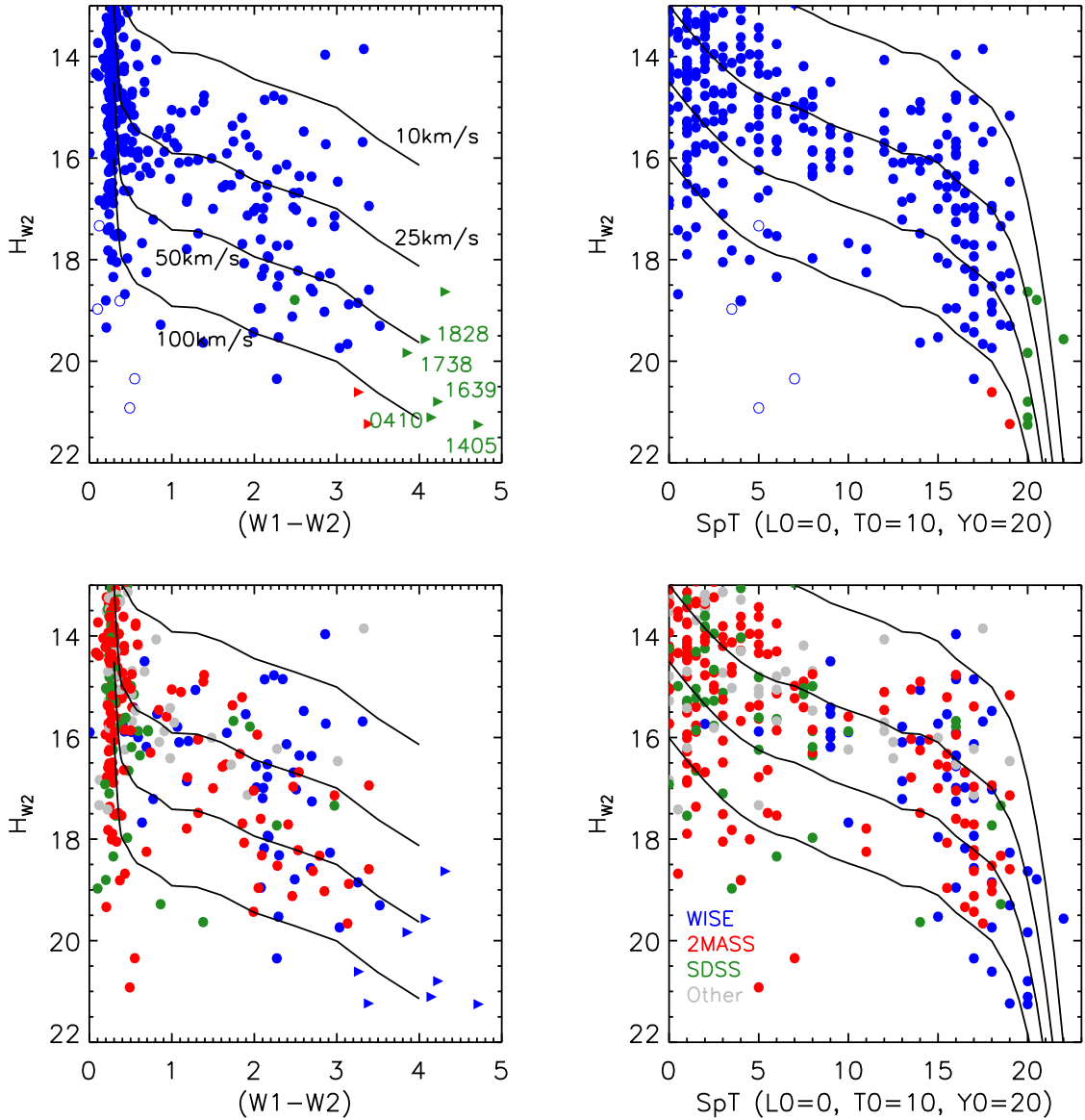}
\caption{Diagrams showing reduced proper motion against colour (left) and spectral type (right). In the top two plots the DwarfArchive and K12 census are plotted in blue using proper motions from K11. Y dwarfs are green and ultracool sub-dwarfs as open symbols. WISE 0013+0634 and WISE 0833+0052 are red. In the bottom two plots the colour coding shows the survey used to identify each object (WISE, 2MASS, SDSS, and others).\label{fig:rpmd}}
\end{center}
\end{figure*}

In Figure \ref{fig:rpmd} we plot reduced proper motion (H$_{W2}$=$W2$+5$\log{\mu}$+5) against $W1-W2$ and spectral type (left and right hand plots respectively). In the top two plots WISE 0013+0634 and WISE 0833+0052 are shown as a red symbols. We also plot (as blue symbols) a sample that includes the WISE census (with proper motions from K11) and all additional L and T dwarfs from DwarfArchives (if they are detected in WISE and have proper motion accuracy $\le$25\%). L sub-dwarfs are shown as open symbols in the top plots, and Y dwarfs are green. Y dwarfs with parallax measurements are individually labeled in the left hand plot. We also over-plot indicative tracks to show characteristic kinematics, for L0-T9.5 and L0-Y2 dwarfs in the left and right plots respectively. These tracks were constructed using the K12 M$_{W2}$ relationship (the fit excluding WISE 1828+2650) for T7-Y2 dwarfs, combined with estimates of M$_{W2}$ for the L0-T6.5 dwarfs (which were derived using an $H-W2$ colour-corrected M$_H$ relation from \citet{marocco2010}). For the left hand plot we made use of a $W1-W2$ colour-spectral type relationship constructed using data from K12.

In the bottom two plots we show the same objects but with different colours representing the surveys in which they were identified. The WISE census is shown in blue, 2MASS in red and SDSS in green. Other objects (grey) mainly come from Denis, with the UKIDSS and CFBDS surveys only yielding 9 and 1 object respectively (that were detected in WISE and with $<25\%$ proper motion uncertainties).

The left hand plots suffer from a bunching-up of the tracks for blue objects due to the relative insensitivity of the $W1-W2$ colour to types earlier than $\sim$mid L. However, the right hand plots shows a similar effect for the latest objects, as the M$_{W2}$ plummets in the Y dwarf regime. For the bluest colours (around L0) there is a full spread across the V$_{tan}$=10-100km~s$^{-1}$ range. This is the most populated part of the diagram. Moving to redder, later types the diagram becomes more sparsely populated, and it is clear that L-mid T dwarfs ($W1-W2<$2) with V$_{tan}$=50-100km~s$^{-1}$ are rare compared to those with V$_{tan}$=10-50km~s$^{-1}$. This is consistent with previous kinematic studies of ultracool dwarfs across the M7-T range \citep[e.g. Figure 7 of ][]{faherty2009} that show a V$_{tan}$ distribution peaking at 10-30 km~s$^{-1}$, with $\sim$85\% having V$_{tan}<$50km~s$^{-1}$ and relatively very few with V$_{tan}\sim$100km~s$^{-1}$ or more. However, for mid T and later ($W1-W2>$2) there appears to be a significantly larger scatter in V$_{tan}$, with WISE 0013+0634 and WISE 0833+0052 establishing the extremes of reduced proper motion for the latest T dwarfs. This may suggest that the fraction of field brown dwarfs with thick-disk/halo kinematics is greater for mid-late T dwarfs than for earlier objects. Indeed, the start of this trend may be evident in figure 6 of \citet{faherty2009}.

The situation is slightly more ambiguous for the Y dwarfs, since the tracks only extend out to T9.5 colours in the left hand plot, and in the right hand plot they steepen rapidly into the Y dwarf regime. However, parallax distance estimates are available for five of the Y dwarfs in the diagram, and are suggestive. WISE 1828+2650 and WISE 1738+2732 have V$_{tan}$ in the range 25-95km~s$^{-1}$, while WISE 0410+1502 and WISE 1405+5534 have V$_{tan}>$100km~s$^{-1}$, and WISE 1639-6847 has V$_{tan}$=73$\pm$8km~s$^{-1}$ \citep[see also ][]{tinney2012}.

The bottom two plots show that the larger scatter in mid-late T reduced proper motion is evident amongst the samples from both WISE and 2MASS, though there are insufficient SDSS objects plotted to assess this survey sample. We considered possible selection biases for these survey samples that might contribute to the scatter. \citet{leggett2010} examines the mid-infrared properties of late T dwarfs, and highlights the T6-6.5 dwarfs 2MASS 0937+2931 and 2MASS 1237+6526. These T dwarfs ($T_{\rm eff}$=800-975K) are believed to be metal poor ([M/H]$\simeq$ -0.3 and -0.2 respectively) from spectroscopic fitting \citep{geballe2009,liebert2007}, and show M$_{[4.5]}$ excess (at a level of $\sim$0.5 mags) with respect to the solar/rich metallicity T7.5 dwarfs Gl570D and HD3651B ($T_{\rm eff}$=780-840K). If metal poor mid-late T dwarfs can be detected out to distances $\ge$25\% further than their solar/rich analogues, we might expect a factor $\ge$2 increase in their number in a $W2$-limited WISE sample. Metal poor T dwarfs could constitute a kinematically older population with larger velocity dispersion. Also standard 2MASS T dwarf selection techniques require blue $H-K$ colours which will introduce a bias towards older lower metallicity objects, which could affect their kinematics in a similar sense. To address these possible biases it would be interesting to compare the reduced proper motions of a large sample of mid-late T dwarfs from the UKIDSS and VISTA surveys \citep[e.g. building on ][]{burningham2013,lodieu2012a}, since those selection techniques should not lead to an older metal poor bias. And for the Y dwarfs, the kinematic constraints are determined by parallax distances that need to be improved.

If the large scatter in reduced proper motion for late T and Y dwarfs genuinely results from an increased fraction of objects with V$_{tan}$=50-100 km~s$^{-1}$, this may provide useful constraints on the mass function and evolution of brown dwarfs. A 10 Gyr-old early- to mid-T dwarf with $T_{\rm eff}$=1000-1400 K will have a mass of about 70-80 M$_{Jupiter}$, while a 10 Gyr-old late-T to early-Y dwarf with $T_{\rm eff}$=400-800 K will have a mass of 25-65 M$_{Jupiter}$ \citep{burrows2001}. For a flat or slightly declining mass function we would therefore expect to see more late-T/Y than early-mid T thick disk brown dwarfs.

\section{Summary and future work}
\label{sec:sum}

We have presented a new method for identifying late T and Y dwarf candidates in the WISE database. After an initial selection of sources with at least 8 individual frames measured per band, detections in $W2$-only were selected avoiding directions where extinction $A_v>$0.8. We then developed rejection methods to assess the nature of the $W2$ sources, using the information relating to the multiple measurements and the source profile-fitting. These methods were designed to reject non-point-like, variable, and moving (solar system) objects, and were based on measures of the profile fit photometry, the scatter in the individual $W2$ measurements, and the fraction of individual detections in the full $W2$ frame set. To effectively trace the desired parameter-space we made use of a control sample of isolated non-moving non variable point sources from the SDSS. These methods led to a manageable sample size for visual inspection down to {\tt{w2snr}}=10, and with the additional removal of sources near bright 2MASS stars, down to {\tt{w2snr}}=8.

We identify 158 candidate late objects. For {\tt{w2snr}}$>$10 $\sim$45\% are not identified by the previously published method of K12, and for the lower {\tt{w2snr}}=8-10 sources $\sim$90\% fall outside the K12 criteria. The main reason for this is the differing approaches used to place constraints on the number of individual $W2$ detections. The main source of incompleteness in our sample is source blending, which can affect the statistical properties that we assess, although we do not directly reject blends in our search method.

Our analysis of WISE 1639-6847 shows that the $\sim$6 month baseline offered by the full WISE data-set can yield reasonably accurate proper motions for high proper motion objects, and that difference images may be a useful tool.

We have identified two high proper motion late T dwarfs, a T8 WISE 0013+0634 and a T9 WISE 0833+0052. These objects both show mid-infrared/near-infrared excess and $K-$band suppression consistent with high gravity and/or low metallicity. Distance estimates lead to kinematics that are consistent with thick disk or halo membership, and current constraints suggest that WISE 0833+0052 may have unambiguous halo kinematics.

Examination of reduced proper motion diagrams suggests that late T and Y dwarfs may consist of a larger fraction of thick-disk/halo members than is the case for earlier objects.

Parallax and radial velocity measurements for  WISE 0013+0634 and WISE 0833+0052 are vital to properly determine the space motion of these objects. And additional near-infrared followup will allow us to categorise candidates in our full sample (according to Table \ref{tab:classes}), and build up the known number of very late T and Y dwarfs across the range of kinematic populations.

\section*{Acknowledgments}

This publication makes use of data products from the Wide-field Infrared Survey Explorer, which is a 
joint project of the University of California, Los Angeles, and the Jet Propulsion Laboratory/California 
Institute of Technology, funded by the National Aeronautics and Space Administration.
The UKIDSS project is defined in Lawrence et al. (2007). UKIDSS uses the UKIRT WFCAM (Casali et al. 2007) 
and a photometric system described in Hewett et al. (2006). The CASU pipeline processing and science archive 
are described in Irwin et al. (2004) and Hambly et al. (2008).
Based on observations obtained as part of the VISTA Hemisphere Survey, ESO Programme 179.A-2010 (PI: McMahon).
This paper includes data gathered with the 6.5 meter Magellan Telescopes located at Las Campanas Observatory, 
Chile (program number CN2012B-019).
Based on observations made with the Italian Telescopio Nazionale Galileo (AOT26TAC68/AOT22TAC96) operated 
on the island of La Palma by the Fundaci\'{o}n Galileo Galilei of the INAF (Istituto Nazionale di Astrofisica) 
at the Spanish Observatorio del Roque de los Muchachos of the Instituto de Astrofisica de Canarias.
Observations have been made using the SofI instrument on ESO's New Technology Telescope as part of programme 
090.C-0791(A).
DP, NL, MG and HJ have received support from RoPACS during this research, and JG, ADJ and JJ have been supported by 
RoPACS, a Marie Curie Initial Training Network funded by the European Commission’s Seventh Framework Programme.
ADJ is supported by a Fondecyt Postdoctorado under project number 3100098.
SKL is supported by the Gemini Observatory, which is operated by AURA, on behalf of the international 
Gemini partnership of Argentina, Australia, Brazil, Canada, Chile, the United Kingdom, and the United 
States of America.
MG is financed by the GEMINI-CONICYT Fund, allocated to the project 32110014.
MTR is partially funded by PB06, CATA.
RK acknowledges partial support from FONDECYT through grant number 1130140.
NL is funded by the national program AYA2010-19136 funded by the Spanish ministry of science and innovation.
VJSB has been supportedby the Spanish Ministryof Economics and Competitiveness under the project AYA2010-20535.
This research has made use of the SIMBAD database, operated at CDS, Strasbourg, France.

\bibliographystyle{mn2e}
\bibliography{refs}

\end{document}

%% file: aas_macros.tex
%
%
%
%


\def\aj{\rm{AJ}}                   
\def\araa{\rm{ARA\&A}}             
\def\apj{\rm{ApJ}}                 
\def\apjl{\rm{ApJ}}                
\def\apjs{\rm{ApJS}}               
\def\ao{\rm{Appl.~Opt.}}           
\def\apss{\rm{Ap\&SS}}             
\def\aap{\rm{A\&A}}                
\def\aapr{\rm{A\&A~Rev.}}          
\def\aaps{\rm{A\&AS}}              
\def\azh{\rm{AZh}}                 
\def\baas{\rm{BAAS}}               
\def\jrasc{\rm{JRASC}}             
\def\memras{\rm{MmRAS}}            
\def\mnras{\rm{MNRAS}}             
\def\pra{\rm{Phys.~Rev.~A}}        
\def\prb{\rm{Phys.~Rev.~B}}        
\def\prc{\rm{Phys.~Rev.~C}}        
\def\prd{\rm{Phys.~Rev.~D}}        
\def\pre{\rm{Phys.~Rev.~E}}        
\def\prl{\rm{Phys.~Rev.~Lett.}}    
\def\pasp{\rm{PASP}}               
\def\pasj{\rm{PASJ}}               
\def\qjras{\rm{QJRAS}}             
\def\skytel{\rm{S\&T}}             
\def\solphys{\rm{Sol.~Phys.}}      
\def\sovast{\rm{Soviet~Ast.}}      
\def\ssr{\rm{Space~Sci.~Rev.}}     
\def\zap{\rm{ZAp}}                 
\def\nat{\rm{Nature}}              
\def\iaucirc{\rm{IAU~Circ.}}       
\def\aplett{\rm{Astrophys.~Lett.}} 
\def\apspr{\rm{Astrophys.~Space~Phys.~Res.}}
\def\bain{\rm{Bull.~Astron.~Inst.~Netherlands}} 
\def\fcp{\rm{Fund.~Cosmic~Phys.}}  
\def\gca{\rm{Geochim.~Cosmochim.~Acta}}   
\def\grl{\rm{Geophys.~Res.~Lett.}} 
\def\jcp{\rm{J.~Chem.~Phys.}}      
\def\jgr{\rm{J.~Geophys.~Res.}}    
\def\jqsrt{\rm{J.~Quant.~Spec.~Radiat.~Transf.}}
\def\memsai{\rm{Mem.~Soc.~Astron.~Italiana}}
\def\nphysa{\rm{Nucl.~Phys.~A}}   
\def\physrep{\rm{Phys.~Rep.}}   
\def\physscr{\rm{Phys.~Scr}}   
\def\planss{\rm{Planet.~Space~Sci.}}   
\def\procspie{\rm{Proc.~SPIE}}   

\let\astap=\aap
\let\apjlett=\apjl
\let\apjsupp=\apjs
\let\applopt=\ao